\documentclass[letterpaper, 11pt]{article}
\pdfoutput=1

\usepackage[nocompress]{cite}
\usepackage{jheppub}

\usepackage{graphicx}
\usepackage{epstopdf}
\usepackage{amsmath, amssymb}

\usepackage{bm}
\usepackage{bbold}

\usepackage[us,12hr]{datetime} 

\newcommand{\be}{\begin{eqnarray}}
\newcommand{\ee}{\end{eqnarray}}
\newcommand{\nn}{\nonumber}
\newcommand{\bn}{\begin{enumerate}}
\newcommand{\en}{\end{enumerate}}

\def\IC{\mathbb{C}}

\def\IO{\mathbb{O}}

\def\IR{\mathbb{R}}
\def\IZ{\mathbb{Z}}

\def\CH{{\cal H}}
\def\CI{{\cal I}}

\def\CM{{\cal M}}
\def\CN{{\cal N}}
\def\CO{{\cal O}}

\def\CS{{\cal S}}

\def\CV{{\cal V}}
\def\CW{{\cal W}}

\def\a{\alpha}

\def\s{\sigma}

\def\G{\Gamma}

\def\half{\frac{1}{2}}

\newcommand{\ket}[1]{|{#1}\rangle}

\newtheorem{Conjecture}{Conjecture}

\def\Tr{{\rm Tr}}
\def\tr{{\rm tr}}

\def\fg{\mathfrak{g}}
\def\PE{\mathrm{PE}}

\def\vec#1{\bm{#1}}
\def\FQ{\mathfrak{Q}}

\title{Vertex operator algebras of Argyres-Douglas theories from M5-branes}

\author[a,b]{Jaewon Song,}
\author[c,d]{Dan Xie,}
\author[e,c,d]{and Wenbin Yan\footnote{Primary affiliation: Yau Mathematical Sciences Center, Tsinghua University, Beijing, China}}

\affiliation[a]{Department of Physics, University of California, San Diego, La Jolla, CA 92093, USA}
\affiliation[b]{School of Physics, Korea Institute for Advanced Study, Seoul 02455, Korea}
\affiliation[c]{Center of Mathematical Sciences and Applications, Harvard University, Cambridge, MA 02138, USA}
\affiliation[d]{Jefferson Physical Laboratory, Harvard University, Cambridge, MA 02138, USA}
\affiliation[e]{Yau Mathematical Sciences Center, Tsinghua University, Haidian District, Beijing, 100084, China}

\emailAdd{jsong@kias.re.kr}
\emailAdd{dxie@cmsa.fas.harvard.edu}
\emailAdd{wbyan@cmsa.fas.harvard.edu}

\preprint{KIAS-P17032}

\abstract
{
We study aspects of the vertex operator algebra (VOA) corresponding to Argyres-Douglas (AD) theories engineered using the 6d $\CN=(2, 0)$ theory of type $J$ on a punctured sphere. 
We denote the AD theories as $(J^b[k],Y)$, where $J^b[k]$ and $Y$ represent an irregular and a regular singularity respectively. We restrict to the `minimal' case where $J^b[k]$ has no associated mass parameters, and the theory does not admit any exactly marginal deformations.
The VOA corresponding to the AD theory is conjectured to be the W-algebra $\CW^{k_{2d}}(J,Y)$, where $k_{2d}=-h+ \frac{b}{b+k}$ with $h$ being the dual Coxeter number of $J$. 
We verify this conjecture by showing that the Schur index of the AD theory is identical to the vacuum character of the corresponding VOA, and the Hall-Littlewood index computes the Hilbert series of the Higgs branch. We also find that the Schur and Hall-Littlewood index for the AD theory can be written in a simple closed form for $b=h$.
We also test the conjecture that the associated variety of such VOA is identical to the Higgs branch. The M5-brane construction of these theories and the corresponding TQFT structure of the index play a crucial role in our computations. 
}

\begin{document}
\maketitle

\section{Introduction}
It has been found recently that for any 4d $\CN=2$ SCFT, there is a protected sector described by a 2d chiral algebra or more precisely a vertex operator algebra (VOA) \cite{Beem:2013sza}. The VOA corresponding to 4d $\CN=2$ SCFT includes the operators in the Higgs branch chiral ring and more generally the so-called Schur operators \cite{Gadde:2011ik,Gadde:2011uv}. Correlation functions in this sector are meromorphic and do not change under exactly marginal deformations. 

The 2d chiral algebra\footnote{We use the vertex operator algebra and chiral algebra interchangeably.} is constructed from the 4d theory as follows. First, pick a two-dimensional slice in the four-dimensional space $\IR^2 \subset \IR^4$ with complex coordinates $(z, \bar{z})$. Then choose a set of particular operators $\CO(z, \bar{z})$ living on this plane $\IR^2$ annihilated by a combination of Poincare and conformal supercharges $\FQ = Q + S$. At the origin these operators are the Schur operators. The operator product expansion (OPE) of these operators turns out to be meromorphic up to the $\FQ$-exact piece. Therefore by passing through the $\FQ$-cohomology, the operators $\CO(z, \bar{z})$ form a meromorphic chiral algebra or VOA. 

Under this 4d $\CN=2$ SCFT/VOA correspondence, the 2d Virasoro central charge is given in terms of 4d central charge as 
\begin{align}
 c_{2d} = -12 c_{4d} \ .  
\end{align}
When there is a (non-R) global symmetry $\fg$, the symmetry in 2d is enhanced to affine symmetry $\hat{\fg}_{k_{2d}}$ with the level given by the 4d flavor central charge as
\begin{align}
 k_{2d} = - \half k_{4d} \ . 
\end{align}
The (super)-character of the VOA gives the Schur limit of the superconformal index
\begin{align}
 \Tr_{\CV} (-1)^F q^{L_0} = \CI_{\mathrm{Schur}}(q) \ , 
\end{align}
where $\CV$ denotes the VOA or the vacuum module of the corresponding chiral algebra. 
We review the definition of the superconformal index and its various limits in appendix \ref{app:n2idx}. 

Various aspects of the VOA corresponding to the 4d $\CN=2$ SCFT has been studied. The topological field theory (TQFT) structure of the chiral algebra for the class $\CS$ theory has been investigated in \cite{Beem:2014rza, Lemos:2014lua}. The chiral algebra for the (generalized) Argyres-Douglas theory \cite{Cordova:2015nma, Xie:2016evu, Creutzig:2017qyf} and also for the $\CN=3$ theories \cite{Nishinaka:2016hbw, Lemos:2016xke} has been identified. Bounds on the central charges have been obtained using the chiral algebra structure \cite{Liendo:2015ofa, Lemos:2015orc}. The effect of the defect operators has been studied in \cite{Cordova:2016uwk, Cordova:2017mhb}. 
It has also been conjectured that it is possible to reconstruct the Higgs branch \cite{BRVOA} and also Macdonald index from the chiral algebra \cite{Song:2016yfd}. Moreover, the Coulomb branch index is related to the modular transformation of the VOA \cite{Fredrickson:2017yka}.


Our primary focus in this paper is to understand the Schur sector and the Higgs branch of the generalized Argyres-Douglas theory \cite{Argyres:1995jj,Argyres:1995xn} that can be engineered from M5-branes \cite{Gaiotto:2009hg,Gaiotto:2009we, Xie:2012hs, Wang:2015mra}. 
They are interesting because the corresponding chiral algebra/VOA for a subset of such AD theories are particularly simple \cite{Xie:2016evu}. In this paper, we focus on the subset of the AD theories, where 1) there is no mass parameters associated to the irregular singularity $J^b[k]$, and 2) there is no exactly marginal deformations. It is widely believed that every exactly marginal operator in $4d$ $\CN=2$ SCFT comes from the gauge coupling. Therefore the existence of a marginal coupling signals that this theory can be decomposed into decoupled SCFTs by setting the gauge coupling to zero. We will focus on the ``non-decomposable" AD theories.\footnote{For example, $D_4^6[3] \equiv (D_4^6[3], \varnothing) $ theory has an exactly marginal operator. This theory can be decomposed into three copies of the $(A_1^2[1], F)$ theory (also called the $H_1 = (A_1, A_3)$ theory in the literatures) coupled via $SU(2)$ gauge group. The associated VOA is conjectured in \cite{Buican:2016arp}.} 

The main tools we use are the conjectural vertex operator algebra and the topological field theory structure suggested by 
the M5-brane construction of the class $\CS$ theory theories. The Schur index for any class $\CS$ theory can be obtained by using the 4d/2d correspondence \cite{Gadde:2009kb,Gadde:2011ik,Gadde:2011uv,Gaiotto:2012xa} between 4d SCFT and 2d topological quantum field theory (TQFT). This correspondence has also been extended to the case with irregular punctures \cite{Buican:2015ina,Buican:2015tda,Song:2015wta, Buican:2017uka}, which allows us to obtain the index for the AD theory. It has also been found to be consistent with the vacuum character of the chiral algebra for the AD theory. The result also agrees with the IR computation in the Coulomb branch \cite{Cordova:2015nma, Cecotti:2015lab, Cordova:2016uwk, Cordova:2017ohl} done by using the BPS monodromy operator \cite{Cecotti:2010fi,Iqbal:2012xm}. The full superconformal index has been computed using the renormalization group flow from certain $\CN=1$ gauge theory to the $\CN=2$ AD theory \cite{Maruyoshi:2016tqk, Maruyoshi:2016aim, Agarwal:2016pjo}, which gives a consistent result with the chiral algebra. 


We consider a class of AD theories labeled as $(J^b[k], Y)$.\footnote{When $b=h^\vee$ and $Y=F$, this also is identical to the $D_p(J)$ theory of \cite{Cecotti:2012jx,Cecotti:2013lda} with some value $p$.} This class of SCFTs is engineered using 6d $\CN=(2,0)$ theory of type $J \in ADE$ on a Riemann sphere with one irregular singularity labeled by $J^b[k]$, and one regular 
singularity labeled by $Y$ which also specifies a nilpotent orbit\footnote{In this paper, we use the Higgs branch label for a puncture, i.e. a full regular puncture corresponds to a trivial nilpotent orbit, while null regular puncture corresponds to 
the principal nilpotent orbit.} of $J$. When there is no regular puncture (or $Y=\varnothing$), we use the notation $J^b[k]\equiv(J^b[k], \varnothing)$ to denote the corresponding theory. 

When the regular puncture is full ($Y=F$) or null ($Y=\varnothing$), the VOAs for the corresponding theories $(J^b[k], F)$ and $J^b[k]$ have been conjectured in \cite{Xie:2016evu}. On the other-hand, in \cite{Beem:2014rza} it was shown that the associated VOA for the theories given by other type of punctures can be obtained via quantum Drinfeld-Sokolov reduction \cite{Drinfeld:1984qv, Bershadsky:1989mf,Bershadsky:1989tc,Feigin:1990pn}. Combining the two, we obtain the following:
\begin{itemize}
\item
The VOA for the $(J^b[k], Y)$ theory is given by the W-algebra $\CW^{k_{2d}}(J, Y)$, 
where $k_{2d}=-h+{b\over b+k}$ with $h$ being the dual Coxeter number of $J$. Here $\CW^{k_{2d}}(J, Y)$ is the W-algebra obtained via quantum Drinfeld-Sokolov reduction  of the Kac-Moody algebra $\widehat{J}_{k_{2d}}$ using the nilpotent orbit associated to $Y$:
\begin{align}
 \widehat{J}_{k_{2d}}= \CW^{k_{2d}}(J, F) \leadsto \CW^{k_{2d}}(J, Y)
\end{align}
Here $F$ denotes the full puncture carrying the flavor symmetry $J$. 
\end{itemize}
As was conjectured in \cite{Beem:2014rza}, the VOA also allows us to extract the Higgs branch of the 4d theory. Therefore we have:
\begin{itemize}
\item
The Higgs branch of theory $(J^b[k], Y)$ is identified with the associated variety $X_\CV$ \cite{MR3456698} of $\CW^{k_{2d}}(J, Y)$:\footnote{This relation is not strictly true for a general 4d $\CN=2$ SCFT, but we provide evidence for the examples we consider in this paper. See \cite{Beem:2014rza} for more detail.}
\begin{align}
\boxed{ \CM_{\mathrm{Higgs}} = X_\CV }
\end{align}
The associated variety $X_\CV$ is given in terms of  the closure of a nilpotent orbit, which depends on the choice of $k$ and $Y$ as:
\begin{itemize}
\item If $k>0$, the associated variety is 
\begin{align}
X_{\CV}= N \cap S_Y \ , 
\end{align}
where $N$ is the principal nilpotent cone, and $S_Y$ is the Slodowy slice defined using the nilpotent orbit corresponding to $Y$.
\item If $k<0$, the associated variety is given as
\begin{equation}
X_{\CV}=X_M \cap S_Y \ , 
\end{equation}
where $X_M = \overline{\mathbb{O}[k]}$ is the closure of a certain nilpotent orbit which \emph{depends on $k$}.
\end{itemize}
\end{itemize}

The goal of the current paper is to verify the conjectures above. We test them by computing the Schur and Hall-Littlewood indices of the $(J^b[k], Y)$ theory and comparing them with the character and the associated variety \cite{BRVOA,MR3456698} of the chiral algebra $\CW^{k_{2d}}(J, Y)$. We also utilize the 3d mirror symmetry whenever it is available. 
One of the key result of the current paper is the following:
\begin{itemize}
\item 
The Schur index of the theory $(J^b[k], Y)$ is given by the vacuum character of algebra $\CW^{k_{2d}}(J, Y)$. In particular, when $b=h$ and $Y=F$ (the full puncture), the character has a surprisingly simple formula in terms of plethystic exponential,
\begin{align}\label{eq:IdxFull}
\boxed{
 \CI_{(J^h[k], F)} (q; \vec{z}) = \PE \left[ \frac{q - q^{h+k}}{(1-q)(1-q^{h+k})}\chi^F_{\textrm{adj}} (\vec{z}) \right]  
}   \ . 
\end{align}
When there is no regular puncture, or equivalently $Y=\varnothing$, the Schur index is given by
\begin{align}
\boxed{
  \CI_{J^h [k]}(q)  = \frac{1}{\prod_{i=1}^r (q^{d_i}; q)} \PE \left[ - \frac{q^{h+k}}{1-q^{h+k}} \chi_{\textrm{adj}} (q^{\vec\rho}) \right] 
} \ . 
\end{align}
Here $d_i$ are the degrees of the Casimirs of $J$ and $\vec{\rho}$ is the Weyl vector. The plethystic exponential is defined as 
\begin{equation}
\PE\left[x\right] = \exp\left(\sum_{n=1}^{\infty}\frac{1}{n}x^n\right).
\end{equation}
We also give a similar expression for the general puncture of type $Y$ in section \ref{sec:Schur}. 
\end{itemize}
It was proven that our formula for the Schur index \eqref{eq:IdxFull} for the $(J_h[k], F)$ theory is indeed identical to the vacuum character of the corresponding VOA \cite{kac2017remark}. 

This paper is organized as follows. In section \ref{sec:ADthy}, we review some basic facts about the AD theories we consider. In section \ref{sec:ADVOA}, we describe the corresponding
vertex operator algebras of our theories. In section \ref{sec:Schur}, we describe the TQFT approach to the Schur index and thereby giving a physical derivation of the character formula for the associated chiral algebra. In section \ref{sec:Higgs}, we study the Higgs branch from three perspectives:  
the associated variety of the vertex operator algebra, the 3d mirror symmetry, and the TQFT approach. We find these different approaches give consistent results. Finally, we conclude with a remark in section \ref{sec:Conclusion}. In the appendix, we review the definition of the superconformal index and its limits and also list the nilpotent orbits which can appear as the Higgs branch of the AD theories considered in this paper. 

\section{Generalized Argyres-Douglas theories from M5-branes} \label{sec:ADthy}

One can engineer four-dimensional $\mathcal{N}=2$ Argyres-Douglas SCFT by putting six dimensional $\CN=(2,0)$ theory of type $J$ on a two-dimensional Riemann surface with the following configurations: a) Sphere with an
irregular singularity; b) Sphere with one irregular singularity and one regular singularity.  
The regular singularity is classified in terms of the 
nilpotent orbits\footnote{We use the Higgs branch label, i.e. a trivial nilpotent orbit represents a full puncture.} \cite{Gaiotto:2009hg,Chacaltana:2012zy} of the Lie algebra $J$. 
 The classification of the irregular singularity is related to the classification of positive grading of Lie algebra $J$ \cite{Xie:2012hs, Wang:2015mra, Xie:2017vaf}.

\subsection{Irreducible irregular singularity}
\textbf{Irregular singularity} Let us start with the 6d $\CN=(2,0)$ theory of type $J$ and compactify it on a Riemann surface $\Sigma$,
to get a four dimensional $\mathcal{N}=2$ theory. The Coulomb branch of 
the 4d theory is described by a Hitchin system defined on $\Sigma$. 
The Hitchin system involves a holomorphic one-form $\Phi$ which is called the Higgs field, 
and the irregular singularity is defined by the following singular boundary condition:
\begin{equation}
\Phi={T_k\over z^{2+{k\over b}}}+\ldots\,.
\end{equation}
Here  $T_k$ is a regular semi-simple element of $J$. The allowed value of $b$ has been classified in \cite{Wang:2015mra} and summarized in table \ref{table:isolatedsingularitiesALEfib}. We label these irregular singularities as $J^b[k]$.  The mass parameters are 
identified with the parameters appear as the coefficient of the first order pole. We list the type of irregular singularities that do not admit any mass parameter in table \ref{table:constraint}.  
We call them as irreducible irregular singularities.
One can get an AD theory using a single irreducible irregular singularity of the above type, and the central charge for such theory is shown in table \ref{table:noflavor:centralcharges}. 
One can also find the Coulomb branch spectrum by studying the Seiberg-Witten geometry.
\begin{table}[h]
\begin{center}
  \begin{tabular}{ |c|c|c|c| }
    \hline
       $J$& Singularity & $b$ &$\mu_0$  \\ \hline
     $A_{N-1}$ &$x_1^2+x_2^2+x_3^N+z^k=0$&  $N$& $(N-1)(k-1)$\\ \hline
 $~$& $x_1^2+x_2^2+x_3^N+x_3 z^k=0$ & $N-1$& $N(k-1)+1$\\ \hline

 $D_N$   & $x_1^2+x_2^{N-1}+x_2x_3^2+z^k=0$ & $2N-2$ &$N(k-1)$ \\     \hline
  $~$   &$x_1^2+x_2^{N-1}+x_2x_3^2+z^k x_3=0$& $N$& $2k(N-1)-N$ \\     \hline

  $E_6$  & $x_1^2+x_2^3+x_3^4+z^k=0$&12 & $6k-6$  \\     \hline
   $~$  & $x_1^2+x_2^3+x_3^4+z^k x_3=0$ &9& $8k-6$   \\     \hline
  $~$  & $x_1^2+x_2^3+x_3^4+z^k x_2=0$  &8& $9k-6$   \\     \hline

   $E_7$  & $x_1^2+x_2^3+x_2x_3^3+z^k=0$& 18 & $7k-7$  \\     \hline
      $~$  & $x_1^2+x_2^3+x_2x_3^3+z^kx_3=0$ &14& $9k-7$    \\     \hline

    $E_8$   & $x_1^2+x_2^3+x_3^5+z^k=0$&30 & $8k-8$  \\     \hline
        $~$   & $x_1^2+x_2^3+x_3^5+z^k x_3=0$ &24& $10k-8$   \\     \hline
    $~$   & $x_1^2+x_2^3+x_3^5+z^k x_2=0$ & 20 &$12k-8$ \\     \hline
  \end{tabular}
  \end{center}
  \caption{3-fold singularities corresponding to our irregular punctures \cite{Wang:2015mra}, where $\mu_0$ is the dimension of charge lattice. When $b=h$ with h the dual Coxeter number, such 
  theories are also called $(J, A_{k-1})$ which were first studied in \cite{Cecotti:2010fi}.  }
  \label{table:isolatedsingularitiesALEfib}
\end{table}
\begin{table}[h]
\begin{center}
\begin{tabular}{|c|c|c|c|}
\hline
  ${\cal T}$ &$~$&${\cal T}$&$~$  \\ \hline
     $A_{N-1}^N[k]$ &$(k,N)=1$& $A_{N-1}^{N-1}[k]$ &$\text{No solution}$\\ \hline
          $D_{N}^{2N-2}[k]$ &$k\neq 2n$& $D_{N}^{N}[k]$&$N=2^m(2i+1),k\neq 2^m n$\\ \hline
     $E_{6}^{12}[k]$ &$k\neq 3n$& $E_6^{9}[k]$ &$k\neq 9n$\\ \hline
     $E_{6}^8[k]$ &$\text{No solution}$& $E_{7}^{18}[k]$ &$k\neq 2n$\\ \hline
     $E_7^{14}[k]$ &$k\neq 2n,n>1$& $E_{8}^{30}[k]$ &$k\neq 30n$\\ \hline
     $E_{8}^{24}[k]$ &$k\neq 24n$& $E_{8}^{20}[k]$ &$k\neq 20 n$ \\ \hline
\end{tabular}
\end{center}
\caption{Constraint on $k$ so that the irregular singularity $J^b[k]$ has no flavor symmetry. Here $n \in \IZ$.}
  \label{table:constraint}
\end{table}
\begin{table}[h]
\begin{center}
  \begin{tabular}{ |c|c|c|c| }
    \hline
  ${\cal T}$ &$c_{4d}$&${\cal T}$&$c_{4d}$  \\ \hline
     $A_{N-1}^N[k]$ &${(N-1)(k-1)(N+k+Nk)\over 12(N+k)}$& $A_{N-1}^{N-1}[k]$ &${(Nk-N+1)(N+k+Nk-1)\over 12(N-1+k)}$\\ \hline
          $D_{N}^{2N-2}[k]$ &${N(k-1)(-2-k+2N+2kN)\over 12(-2+k+2N)}$& $D_{N}^{N}[k]$ &${((N-1)2k-N)(N+k(2N-1))\over 12(k+N)}$\\ \hline
     $E_{6}^{12}[k]$ &${(k-1)(12+13k)\over 2(12+k)}$& $E_6^{9}[k]$ &${(4k-3)(13k+9)\over 6(9+k)}$\\ \hline
     $E_{6}^8[k]$ &${(3k-2)(13k+8)\over4(8+k)}$& $E_{7}^{18}[k]$ &${7(k-1)(19k+18)\over 12(18+k)}$\\ \hline
     $E_7^{14}[k]$ &${(9k-7)(19k+14)\over 12(14+k)}$& $E_{8}^{30}[k]$ &${2(k-1)(30+31k)\over 3(30+k)}$\\ \hline
     $E_{8}^{24}[k]$ &${(5k-4)(24+31k)\over 6(24+k)}$& $E_{8}^{20}[k]$ &${(3k-2)(20+31k)\over3(20+k)}$\\ \hline
       \end{tabular}
  \end{center}
  \caption{Central charges for the AD theories without flavor symmetries. Note that this formula is only true for certain choice of $k$ and $N$ so that the irregular puncture is irreducible.}
  \label{table:noflavor:centralcharges}
\end{table}

\newpage
\noindent \textbf{Regular singularity} One can also add a regular singularity on top of the irregular singularity to obtain Argyres-Douglas theories carrying a non-abelian flavor symmetry. 
The regular singularity is classified by the embeddings $\Lambda_Y$ of $SU(2)$ into $J$, or equivalently by the nilpotent orbits of the Lie algebra $J$. The flavor symmetry associated to a puncture of type $Y$ is given by the commutant of $\Lambda_Y(\s^+)$ in $J$. 
We list some special orbits in table \ref{special}. 
\begin{table}[h]
\begin{center}
\begin{tabular}{|c|c|c|}
\hline
 Nilpotent orbit & Dimension & Flavor symmetry  \\ \hline
     Maximal (Principal) & $\textrm{dim}J-\textrm{rank}(J)$ & N/A \\ \hline  
         Sub-regular &$ \textrm{dim}J-\textrm{rank}(J)-2$& ~ \\ \hline         
        Minimal&  $d_1$ & ~ \\ \hline   
        Trivial & 0 & $J$  \\ \hline
\end{tabular}
\end{center}
\caption{Some special regular singularities. 
Here $d_1$ is equal to one plus the number of positive roots not orthogonal to the highest root. }
  \label{special}
\end{table}

We label our theory as $(J^b[k], Y)$ Here $Y$ represents the regular puncture (or its corresponding nilpotent orbit) and $J^b[k]$ denotes the irreducible irregular puncture (has no mass parameter). We often denote $F$ as the full (regular) puncture and $\varnothing$ as the null (or the absense of) puncture. We write the central charges\footnote{Here we use the normalized the flavor central charge $k_F$ so that a free hypermultiplet has $k_{SU(2)}=1$.} for such theories in table \ref{table:centralcharge:ADmatter}. 

For $J=A_{N-1}$, the regular singularities are classified by the partitions of $N$ or Young Tableaux $Y=[n_1^{h_1},\ldots, n_s^{h_s}]$ with $\sum_i h_i n_i = N$, and the 
corresponding flavor symmetry is given as
\begin{equation}
G_Y=\left(\prod_{i=1}^s U(h_i)\right)/ U(1).
\end{equation}
For the classification of the regular punctures and the corresponding flavor symmetries of other Lie algebras, see \cite{Chacaltana:2012zy}.

\begin{table}[!htb]
\begin{center}
  \begin{tabular}{ |c|c|c|}
    \hline
       Theory &$c_{4d}$&$\half k_F$  \\ \hline
     $(A_{N-1}^N[k],F)$ &$\frac{1}{12}{(N+k-1)(N^2-1)}$& ${N(N+k-1)\over N+k}$\\ \hline
$(A_{N-1}^{N-1}[k],F)$&${(N+1)[N^2+N(k-2)+1]\over 12}$&${(N-1)^2+kN\over N+k-1}$\\ \hline

 $(D_N^{2N-2}[k],F)$   &${1\over 12}N(2N-1)(2N+k-3)$ & ${(2N-2)(2N+k-3)\over 2N-2+k}$ \\     \hline
  $(D_N^{N}[k],F)$  &${(2N-1)[2k(N-1)+N(2N-3)]\over 12}$ & ${2k(N-1)+N(2N-3)\over N+k} $ \\     \hline

 $(E_{6}^{12}[k],F)$ &${13(k+11)\over 2}$  &${12(k+11)\over k+12}$  \\     \hline
 $(E_{6}^{9}[k],F)$& ${13\over 6}(33+4k)$ &$12-{9\over k+9}$   \\     \hline
 $(E_{6}^{8}[k],F)$&${13\over 4}(22+3k)$ &$12-{8\over k+8}$   \\     \hline

$(E_{7}^{18}[k],F)$&${133\over 12}(17+k)$   &${18(k+17)\over k+18}$  \\     \hline
 $(E_7^{14}[k],F)$&${19\over 12}(119+9k)$  &$18-{14\over k+14}$     \\     \hline

 $(E_{8}^{30}[k],F)$ & ${62\over 3} (29+k)$  &${30(k+29)\over k+30}$  \\     \hline
      $(E_{8}^{24}[k],F)$ &${31\over 6} (116+5 k)$   & $30-{24\over k+24}$   \\     \hline
  $(E_{8}^{20}[k],F)$&${31\over3} (58+3 k)$  &$30-{20\over k+20}$ \\     \hline
  \end{tabular}
  \end{center}
  \caption{The central charge $c_{4d}$ and flavor central charge $k_F$ for the Argyres-Douglas matter. }
  \label{table:centralcharge:ADmatter}
\end{table}

\noindent \textbf{Rank 1 SCFTs}  One interesting class of 4d $\CN=2$ SCFTs is the rank 1 SCFTs labeled by $H_0$, $H_1$, $H_2$, $D_4$, $E_6$, $E_7$, $E_8$. They can be realized as the world-volume theory of a single D3-brane probing the 7-brane singularities in F-theory. They have the global symmetry $\varnothing$, $SU(2)$, $SU(3)$, $SO(8)$, $E_6$, $E_7$, $E_8$ respectively. A noticeable feature is that the Higgs branch of each theory is given by the minimal nilpotent orbit of the flavor group, which is the same as the centered 1-instanton moduli space. 
These theories can be realized in our setup as $(A_1^2[3])$,$(A_1^2[1], F)$,$(A_2^3[-1], F)$, $(D_4^4[-3], F)$, $(E_6^9[-8], F)$, $(E_7^{14}[-13], F)$, $(E_8^{24}[-23], F)$ respectively.

\subsection{Theory with 3d Lagrangian mirror}
We can reduce 4d $\mathcal{N}=2$ SCFTs to 3d and flow to IR to obtain 3d $\mathcal{N}=4$ SCFTs.  For a 3d $\mathcal{N}=4$ SCFT A, it 
is often possible to find a mirror SCFT B. The essential feature of 3d mirror symmetry is that the Coulomb branch of theory A is identified with the Higgs branch 
of theory B, and vice versa. Since the 4d Higgs branch is the same as the Higgs branch of the 3d SCFT A obtained via dimensional reduction, we can use the Coulomb branch of the 3d mirror B to study the 4d Higgs branch. For example, if B admits a Lagrangian description, we can compute the Higgs branch index of A through computing the Coulomb branch index of B. 

\begin{figure}[h!]
    \centering
    \includegraphics[width=6.0in]{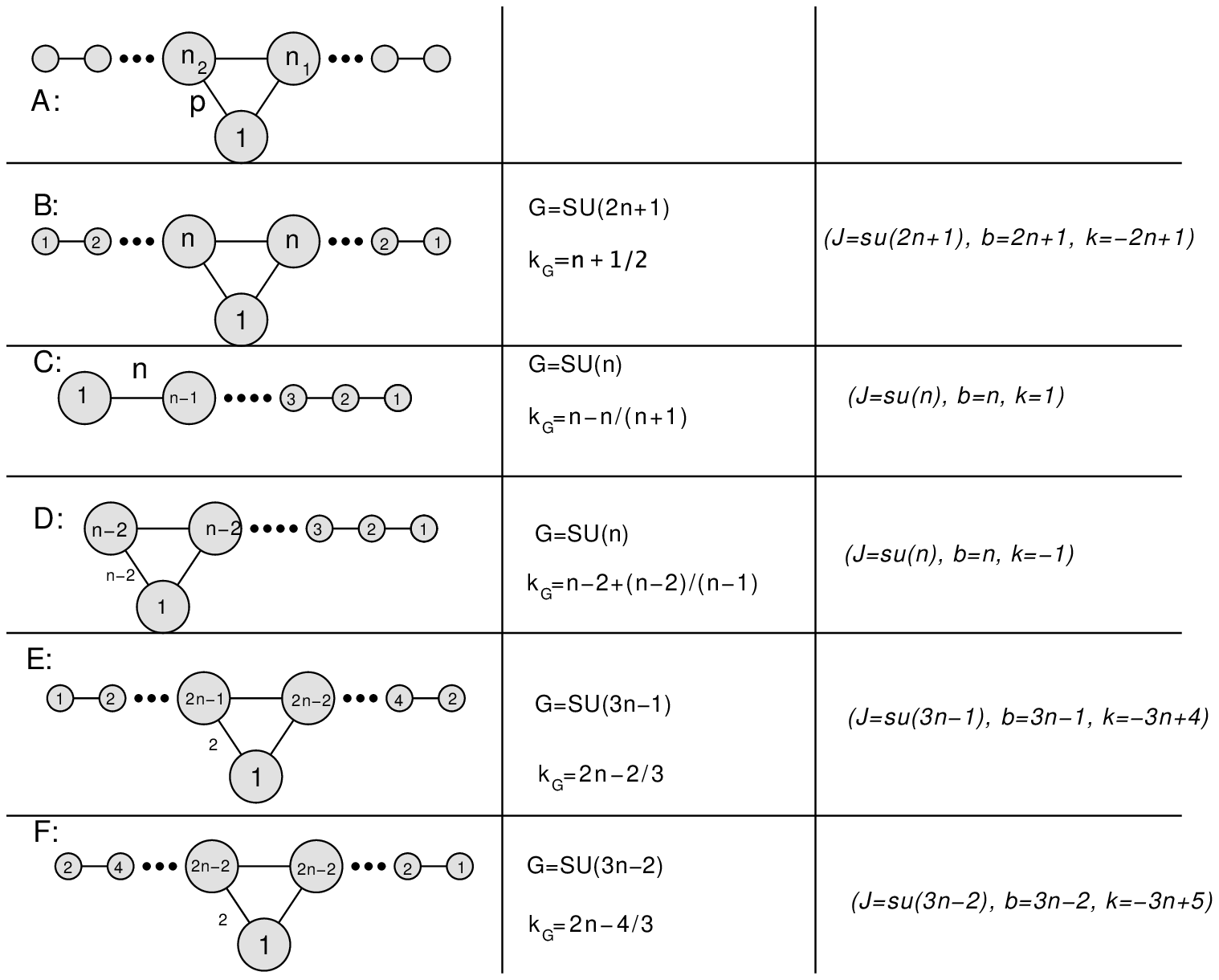}
    \caption{3d Mirror descriptions for some Argyres-Douglas theories considered in this paper, whose flavor symmetry is $SU$. Here, the number $n$ assigned to the edges mean that there are $n$ copies of the bifundamental hypermultiplet. As usual, the overall $U(1)$ factor has to be modded out. 
    $A$ is the general structure of a quiver whose flavor symmetry would be a simple $SU$ group. $B-F$ are examples. 
    We also list the flavor central charge here, and the corresponding theory in terms of the label $(J^b[k],F)$.}
    \label{3dmirror}
\end{figure}
For $J=A_{N-1}$, we find the list of AD theories that admit Lagrangian 3d mirrors, as given in figure \ref{3dmirror}. 
The idea for finding these theories is as follows: first, we use the method of Gaiotto-Witten \cite{Gaiotto:2008ak} to find the 3d $\CN=4$ quivers whose flavor symmetry on the Coulomb branch is $SU(N)$. Then we use the method of \cite{Xie:2012hs} to find a Hitchin system realization where the Higgs field has integral order poles. From the Hitchin system, we can derive the 4d Coulomb branch spectrum. Finally, we search for a realization that fits into the class of theories we label as $(J^b[k],F)$ with $J=A_{N-1}$.

\subsection{Reducible irregular singularity}
In the last subsection, we considered the theory defined by an irregular singularity which does not admit any mass parameter. 
For a general irregular singularity, it was found in \cite{Xie:2017vaf}
that it is useful to represent our theory by an auxiliary punctured sphere.  
Let us consider 6d $A_{N-1}$ $(2,0)$ theory on a sphere with the following irregular singularity 
\begin{align}
\Phi={T\over z^{2+{p\over q}}}+\ldots \ . 
\end{align}
Here $(p,q)$ are co-prime and $T$ is regular semi-simple, and $N=rq$ or $N=rq+1$ with $r$  an integer. One can also add a regular singularity of $A_{N-1}$ type. The range of $(p,q)$ is restricted to  
$q>0,~ p \geq -q +1$. 
Depending on values of $p, q$, it is shown that one 
can have the following representations in terms of auxiliary punctured sphere \cite{Xie:2017vaf}:
\begin{itemize}
\item If $p=0$, we get the usual class ${\cal S}$ theory on a sphere with regular singularity. 
\item If $p=1, q=1$, the theory can be represented by a sphere with $r$ marked points of the black type $Y=[1]$.  
\item If $p>1, q=1$, one can represent the theory by a sphere with $r$ marked points of the black type $Y=[1]$, and one extra red marked point representing the regular singularity. 
\item If $q\neq 1$, one can represent the theory by a sphere with $r$ marked points of the black type, one marked point of the red type representing the regular singularity, and one marked point of the blue type. 
\end{itemize}
For each marked point, one associate a Young Tableaux $Y$ with varying size (except for the class ${\cal S}$ case where the Young Tableaux has the fixed size determined by $J$). We refer to \cite{Xie:2017vaf} for more detail. 

In the previous sections, we have considered the case of $r=1$ and we will be only considering this case in the rest of the current paper (notice that in general this condition implies that the irregular singularity has neither mass parameter nor exact marginal deformation). Therefore our theory can be represented by the following three punctured sphere: one trivial blue marked point, one red marked point representing regular singularity, and one black marked point of type $Y=[1]$. 
We hope that the above representation of more general AD theories and the S-duality proposed in \cite{Xie:2017vaf} can be helpful in studying the indices of those theories.

\section{Vertex operator algebra of the AD theories} \label{sec:ADVOA}
We have introduced a class of AD theories labeled as $(J^b[k], Y)$.  Here $J^b[k]$ represents an irregular singularity without any mass parameter, and $Y$ is an arbitrary 
regular singularity. We also assume that the theory is `indecomposable,' namely has no exactly marginal operators. 

It was shown in \cite{Beem:2013sza} that for any four-dimensional 
$\mathcal{N}=2$ SCFT, one can associate a two-dimensional chiral algebra or vertex operator algebra. The basic correspondence is as follows:
 \begin{itemize}
\item The 2d Virasoro central charge $c_{2d}$ is given in terms of the conformal anomaly $c_{4d}$ of the 4d theory as
\be
	c_{2d} = -12 c_{4d} \ . 
\ee
\item The global symmetry algebra $\mathfrak{g}$ becomes an affine Kac-Moody algebra $\hat{\mathfrak{g}}_{k_{2d}}$ and the level of affine Kac-Moody algebra $k_{2d}$ is given by the 4d the flavor central charge $k_F$ as
\begin{equation}
	k_{2d}=-\half k_F \ . 
\label{eq:centralchargerelation}
\end{equation}
\item The (normalized) vacuum character of the chiral algebra/VOA is identical to the Schur index of the 4d theory:
\begin{align}
 \chi_0(q) = \CI_{\textrm{Schur}}(q) \ . 
\end{align}
\end{itemize}

The VOAs corresponding to the theories $(J^b[k], F)$ or $J^b[k] \equiv (J^b[k], \varnothing)$ with irreducible irregular puncture $J^b[k]$ were conjectured in \cite{Xie:2016evu}. 
Along with the same line as in the case of the usual class $\CS$ theory \cite{Beem:2014rza}, we can further conjecture VOA for the general choice of the puncture $Y$: 
 
\begin{Conjecture}
The VOA corresponding to $(J^b[k], Y)$ is given by the W-algebra $\CW^{k_{2d}}(J, Y)$. 
Here $k_{2d}=-h+{b\over b+k}$ and $\CW^{k_{2d}}(J, Y)$ algebra is the quantum Drinfeld-Sokolov reduction \cite{Drinfeld:1984qv, Bershadsky:1989mf,Bershadsky:1989tc,Feigin:1990pn} of the Kac-Moody algebra 
$\widehat{J}_{k_{2d}}$ using the nilpotent orbit (or the $SU(2)$ embedding into $J$) associated with $Y$. 
\end{Conjecture}
Two extreme situations where $Y$ being $F$(full) or $\varnothing$(null) have been considered in \cite{Xie:2016evu}, and the corresponding VOAs are the affine Kac-Moody algebra $\widehat{J}$ and the standard $W$-algebra associated with $J$. 

For a general choice of $Y$, we obtain the VOA for the corresponding theory via quantum Drinfeld-Sokolov (qDS) reduction. 
The central charge of the vertex operator algebra $\CW^{k}(J,  Y)$ is given as \cite{deBoer:1993iz}
\begin{align}
c_{2d}=\textrm{dim}J_0 -{1\over 2} \textrm{dim} J_{1/2}-{12\over k+h} \big|\rho-(k+h)x \big|^2 \ , 
\label{2dcentral}
\end{align}
where $h$ is the dual Coxeter number, and $\rho$ is the longest root.  
Let us explain the notation: for a nilpotent orbit, one has a $su(2)$ triple $(x,e,f)$ so that $[x,e]=e,~[x,f]=-f, [e,f]=x$. Here $x$ is a semi-simple 
element of the Lie algebra $J$, and $f$ is a nilpotent element. This can be also obtained by choosing a $su(2)$ embedding $\Lambda_Y$ into $J$ given by the choice of $Y$. It gives $(x, e, f) = (\Lambda_Y(\s^3), \Lambda_Y(\s^+), \Lambda_Y(\s^-))$. Now, $J$ has an eigenspace decomposition under the adjoint action of $x$ as:
\begin{align}
J=\bigoplus_{j\in {1\over 2}\IZ} J_j \ , 
\end{align}
Let $J^f$ be the centralizer of $f$ in $J$, which also has a grading by the adjoint action of $x$:
\begin{equation}
J^f= \bigoplus_{j \in \half \IZ} J_{-j}^f 
\end{equation}
For us, $f$ is an element of the nilpotent orbit associated to $Y$. The algebra $\CW^{k_F}(J, Y)$ is strongly generated by $\textrm{dim}J^f$ fields $X^a(z)$, where $a$ runs over 
a basis of $J^f$. The generator $X^a(z)$ has a conformal weight $1+j$ if $X^a\in J_{-j}^f$. 

The level $k_{2d}$ of an affine Kac-Moody algebra $\widehat{J}_{k_{2d}}$ (for $J=ADE$) is called \emph{admissible} 
if it can be written in the following way
\begin{equation}
k_{2d} =-h+{p\over q},~~(p,q)=1,~~p\geq h.
\end{equation}
So only for  $b=h$, the level of our conjectural VOA is admissible. 
The vacuum character of the corresponding affine Kac-Moody algebra with admissible level and the $W$-algebra obtained from qDS are given in \cite{Kac:1988qc} and \cite{Frenkel:1992ju} respectively.

There is one simple consistency check we can do: the 4d central charge can be computed using the method presented in \cite{Xie:2012hs,Xie:2013jc} and can be used to check 
with 2d central charge formula (\ref{2dcentral}). 

\paragraph{Examples}
Let us study several interesting class of examples. 
 $(A_1, ADE)$\footnote{Here $(G_1, G_2)$ means that the corresponding BPS quiver is a product of two Dynkin diagrams for $G_1$ and $G_2$.} theories are a class of well studied Argyes-Douglas theories. 
 The $(A_1, A_{2N})$ theory can be realized by the choice $J=A_1, b=2, k=2N+1$ with $Y=\varnothing$ being trivial. The $(A_1, D_{2N-1})$ theory can be realized from $J=A_1, b=2, k=2N-3$ with $Y=F$ being the full puncture. 
 The $(A_1, E_6)$ theory can be realized from $J=A_2, b=3, k=4$ with $Y=\varnothing$, and the $(A_1, E_8)$ theory can be realized from $J=A_2, b=3, k=5$ with $Y=\varnothing$. The corresponding VOAs are listed in table \ref{A1ADE}.
In the following, we will find VOAs for other AD theories in this class.
 
\textbf{Example 1}: Consider the case with $J=A_{N-1},~ b=N,~k=1$, and $Y=[N-1,1]$.   This theory is identical to the $(A_1, A_{2N-1})$ theory and has $U(1)$ global symmetry (except for $N=2$, which has enhanced $SU(2)$ symmetry). 
 The corresponding VOA is $\CW^{-{N^2\over N+1}}(A_{N-1}, Y)$.

\textbf{Example 2}: Consider the theory defined by data $J=A_{N-1}, ~b=N,~k=-1$ and $Y=[N-2,1,1]$. It gives the $(A_1, D_{2N-2})$ theory which has $SU(2)\times U(1)$ flavor symmetry (except for $N=3$, which has enhanced $SU(3)$ symmetry).
The corresponding VOA is $\CW^{{-N^2+2N\over N-1}}(A_{N-1}, Y)$.

\textbf{Example 3}: Choose $J=A_{N-1}, ~b=N-1,~(k,b)=1$ and $Y$ to be trivial. 
The same theory can be also obtained from $J=A_{N-1},~b=N,~k=N-1$, and a simple puncture $Y=[N-1,1]$. This also belongs to our list. The $(A_1, E_7)$ theory is identical to $(A_2^2[3], [2, 1])$. Using the other description, we find that the corresponding VOA is $\CW^{-{13\over5}}(A_2, [2, 1])$.

The complete results are summarized in table \ref{A1ADE}. We are able to recover what was found in \cite{Creutzig:2017qyf}. 
These theories can also be realized in other ways, for example $(A_1, A_{2N})$ theory can be realized using $J=A_{2N}$. In general, there are many different M5-brane realizations of the identical 4d AD theory. Whenever this happens, we find isomorphisms between W-algebras.

\begin{table}[h]
\begin{center}
\begin{tabular}{|c|c|c|c|}
\hline
 $(G, G')$ & $(J^b[k], Y)$ &VOA&$Y$  \\ \hline
     $(A_1, A_{2N})$ & $(A_1^2 [2N+1], Y) $ &$\CW^{-2+{2\over 2N+3}}(A_1, Y)$& $[2]$ \\ \hline  
      $(A_1, A_{2N-1})$& $(A_{N-1}^N[1], Y)$ & $\CW^{{-N^2\over N+1}}(A_{N-1}, Y)$&$ [N-1,1]$ \\ \hline         
      $(A_1, D_{2N-1})$& $(A_1^2 [2N-3], Y)$ &$\CW^{-2+{2\over 2N-1}}(A_1, Y)$& $[1,1]$\\ \hline      
       $(A_1, D_{2N-2})$& $(A_{N-1}^N[-1], Y)$ &$\CW^{{-N^2+2N\over N-1}}(A_{N-1}, Y)$& $[N-2,1^2]$ \\ \hline  
      $(A_1, E_{6})$&$(A_2^3[4], Y)$  & $\CW^{-{18\over 7}}(A_2, Y)$& $[3]$ \\ \hline         
      $(A_1, E_{7})$&$(A_2^2[3], Y) $ &$\CW^{-{13\over5}}(A_2, Y)$& $[2,1]$ \\ \hline      
      $(A_1, E_{8})$&$(A_2^3[5], Y)$ & $\CW^{-{21\over 8}}(A_2, Y)$& $[3]$ \\ \hline 

\end{tabular}
\end{center}
\caption{The VOA of $(A_1, ADE)$ theories. Here $Y=[n_1,n_2,\ldots, n_s]$ means that the nilpotent orbit has the Jordan block with size $n_1,\ldots, n_s$. }
  \label{A1ADE}
\end{table}

\paragraph{Schur index and the vacuum character}
The Schur index is given by the vacuum character of the corresponding VOA. 
Here, let us consider the case $b=h$ and $(k,h)=1$, then the VOA for the $J^h[k])$ theory is the same as the $W^J[h+k,h]$ minimal model.
The vacuum character takes the following simple form \cite{Kac:1988qc, Frenkel:1992ju, Bouwknegt:1992wg}\footnote{For $J=A_{N-1}$, the character is already considered in \cite{Cordova:2015nma}.}
\begin{equation}\label{eq:chaofminimalmodel}
\chi(q)={1\over \eta(\tau)^r}\sum_{\omega \in \hat{W}} \epsilon(w)q^{{1\over 2 p p'}|p'w(\Lambda^{+}+\rho)-p(\Lambda^{-}+\rho)|^2},
\end{equation}
where $r$ is the rank of the Lie algebra $J$, $\rho$ is the Weyl vector, $\hat{W}$ is the affine Weyl group, and $\epsilon(w)$ is the signature of the affine Weyl group element.
$\Lambda^{+}\in P_{+}^{0}$ and $\Lambda^{-}\in P_{+}^{k}$ are principle admissible weights such that: 
\begin{equation}
{1\over 12}r h(h+1)(p'-p)^2=|p'(\Lambda^{+}+\rho)-p(\Lambda^{-}+\rho)|^2.
\end{equation}
We have the solution: $\Lambda^{+}=0$ and $\Lambda^{-}=(k,0,\ldots,0)$. Substituting into the character formula, we get the Schur index of our 4d theory $J^h[k]$. 

In the next section, we will show that TQFT structure of AD theory dictates the Schur index of $J^h[k]$ theory to be the simple form
\begin{equation}
 \CI_{J^h [k]} (q) = \frac{1}{\prod_{i=1}^r (q^{d_i}; q)} \PE \left[ - \frac{q^{h+k}}{1-q^{h+k}} \chi_{\textrm{adj}} (q^{\vec\rho}) \right] \ , 
\end{equation}
where $d_i$ are the degrees of the Casimirs of $J$. 
We have checked that this formula matches with equation \ref{eq:chaofminimalmodel} up to a high power in $q$.

It is viable to compute the vacuum character of ${\widehat{J}}_{-\half k_F}$ hence the Schur index of $(J^b[k],F)$ by using the Kac-Wakimoto formula. However, we propose a much compact formula for $b=h$ case,
\begin{equation}
 \CI_{(J^h[k], F)} = \PE \left[ \frac{q - q^{h+k}}{(1-q)(1-q^{h+k})}\chi^F_{\textrm{adj}} (\vec{z}) \right] \ . 
\end{equation}
A derivation of this formula is given in section \ref{sec:Schur}.

\section{Schur index and TQFT} \label{sec:Schur}

In this section, we derive a universal formula for the Schur index of the AD theory using the TQFT structure of the index, which provides a strong check for the identification of the corresponding VOA.

\subsection{Wave function for the irregular puncture}
As an intermediate step, we first write the Schur index of the pure Yang-Mills theory in a TQFT form. Even though the superconformal index is properly defined only for a conformal theory, there is increasing evidence that the Schur index makes sense even for a non-conformal theory. See for example \cite{Cordova:2015nma, Cecotti:2015lab, Cordova:2016uwk}.

The Schur index for the pure YM theory of gauge group $J$ is given by
\be
 \CI_{\textrm{SYM}}^{J} = (q; q)^r \oint [d\vec{z}] \prod_{\vec{\a} \in \Delta_{J}} (q \vec{z}^{\vec{\a}}; q)^2 \ , 
\ee
where $r$ is the rank of the gauge group and $\Delta_J$ is the set of all roots. We write the measure factor as
\be
[d\vec{z}] = \frac{1}{|W_J|} \prod_{i=1}^r \frac{dz_i}{ 2\pi i z_i} \prod_{\vec{\a} \in \Delta_J} (1-\vec{z}^{\vec\a}) \ , 
\ee
where $|W_J|$ denotes the order of Weyl group of $J$. Here we used the short-hand notation $\vec{z}^{\vec{\a}} \equiv \prod_i z_i^{\a_i}$. 

The pure YM theory can be obtained from 6d $\CN=(2, 0)$ theory. When $J$ is simply-laced, we pick $J$-type 6d $\CN=(2, 0)$ theory and compactify on a sphere with two identical irregular punctures. The irregular puncture, which we denote as $I_J \equiv {J^{h}[-h+1]}$ (here $h$ is the dual Coxeter number of $J$), realizes a singularity of the form 
\be
 \Phi (z) = \frac{T}{z^{2-\frac{h-1}{h}}} + \cdots = \frac{T}{z^{1+\frac{1}{h}}} + \cdots \ . 
\ee
From the TQFT structure of the index \cite{Gadde:2009kb,Gadde:2011ik,Gadde:2011uv,Gaiotto:2012xa}, the Schur index should be given by
\be \label{eq:SchurYMTQFT}
 \CI_{SYM}^{J} = \sum_{\vec \lambda} \psi_{\vec \lambda}^{I_{J}} \psi_{\vec \lambda}^{I_{J}} \ . 
\ee
The wave function for the puncture is given by \cite{Song:2015wta}
\be \label{eq:SchurYMWaveFtn}
 \psi_{\vec \lambda}^{I_{J}} = \oint [d\vec{z}] \PE \left[ -\frac{q}{1-q} \chi_{\textrm{adj}}(\vec{z}) \right] \chi_{\vec \lambda}(\vec{z}) \ , 
\ee
where $\chi_{\vec \lambda} (\vec{z}) $ is the character (or Schur function) of $J$ for the representation $\vec \lambda$. This is an analog of the Gaiotto state \cite{Gaiotto:2009ma,Keller:2011ek} in the AGT correspondence \cite{Alday:2009aq}.

It is easy to verify that \eqref{eq:SchurYMWaveFtn} reproduces the Schur index for the pure YM. Plugging in the wave function to the RHS of \eqref{eq:SchurYMTQFT}, we get
\begin{align}
\begin{split}
\sum_{\vec \lambda} \psi_{\vec \lambda}^{I_{J}} \psi_{\vec \lambda}^{I_{J}} 
 &= \oint [d\vec{z}][d\vec{z'}] \PE \left[ - \frac{q}{1-q} \left( \chi_{\textrm{adj}} (\vec{z}) + \chi_{\textrm{adj}} (\vec{z'})\right) \right] \chi_{\vec \lambda} (\vec{z}) \chi_{\vec \lambda} (\vec{z'}) \\
 &= \oint [d\vec{z}] \PE \left[ - \frac{2q}{1-q} \chi_{\textrm{adj}}(\vec{z}) \right] = \CI_{SYM}^{J} \ . 
\end{split}
\end{align}
Here we used the relation $\sum_{\vec \lambda} \chi_{\vec \lambda}(\vec{z})\chi_{\vec \lambda}(\vec{z'}) = \frac{1}{\Delta_J(z)}\delta (\vec{z}-\vec{z'})$, where $\Delta_J (z)$ is the Haar measure. 

The wave functions for the $I_{SU(N)}$ are evaluated when $N=2, 3$ \cite{Song:2015wta}. For $N=2$, we get
\begin{align}
 \psi_\lambda^{I_{SU(2)}}(q) 
  &=  \begin{cases} (-1)^{\frac{\lambda}{2}} q^{\half \frac{\lambda}{2} (\frac{\lambda}{2}+1)} & \mbox{$\lambda$ even}, ~~ \\ 
0 & \mbox{$\lambda$ odd}.
 \end{cases} 
\end{align}
When $N=3$, we get
\be
 \psi^{I_{SU(3)}}_{(\lambda_1, \lambda_2)}(q) = 
 \begin{cases}
  q^{k(k+1)+\ell(\ell+1)+k \ell} & \mbox{if $\lambda_1 = 3k, \lambda_2 = 3\ell$} , \\
  -q^{k^2+\ell^2-1+(k-1)(\ell-1)} & \mbox{if $\lambda_1 = 3k-2,  \lambda_2 = 3 \ell-2 $} , \\
  0 & \mbox{otherwise} , 
 \end{cases}
\ee
where $k, \ell \in \IZ_{\ge 0}$. 

The wave function for the irregular puncture $I_{N, k} \equiv A_{N-1}^N[k]$ is conjectured to be given by \cite{Song:2015wta}
\be 
 \psi_{\vec\lambda}^{I_{N, k}} (q) = \psi_{\vec\lambda}^{I_{SU(N)}}(q^{N+k})  \ . 
\ee
Here, we further conjecture that the wave function for the irreducible irregular puncture $J^h[k]$ (with $b=h$) with no flavor symmetry (where the condition is summarized in table \ref{table:constraint}) is given by
\be \label{eq:IrrSchur}
\boxed{ \psi_{\vec \lambda}^{J^h[k]}(q) = \psi_{\vec{\lambda}}^{I_J} (q^{h+k}) } \ . 
\ee
In the following, we compute the Schur indices for a number of examples assuming the relation \eqref{eq:IrrSchur} and find that it agrees with the vacuum character of the VOA. 

\subsection{AD theories of type $J^{h}[k]$}

Let us consider the theory of type $J^{h}[k] \equiv (J^h[k], \varnothing)$, where $h$ is the dual Coxeter number of $J$. As before, we can write the wave function for the irregular puncture realizing the analog of Gaiotto-Whittaker state for the pure Yang-Mills as 
\be
 \psi_{\vec{\lambda}}^{I_J} (q) = \oint [d\vec{z}] \PE \left[ - \frac{q}{1-q} \chi_{\textrm{adj}} (\vec{z}) \right] \chi_{\vec{\lambda}}(\vec{z}) \ , 
\ee
where $\vec{\lambda}$ is a Dynkin label for the representation of $J$. From the TQFT structure, we should have
\be
  \CI_{J^{h} [k]} (q) = \sum_{\vec \lambda} C_{\vec\lambda}^{-1}(q) \psi_{\vec \lambda}^{J^h[k]}(q) 
  = \sum_{\vec \lambda} \frac{\chi_{\vec\lambda}(q^{\vec\rho})}{\prod_{i=1}^r (q^{d_i}; q)} \psi_{\vec \lambda}^{J^h[k]}(q) \ , 
\ee
where $r$ is the rank of group $J$, and $d_i$ are the degrees of the Casimirs of $J$. We conjecture that the wave function for the general irregular puncture is given by
\be
\psi_{\vec \lambda}^{J^h[k]}(q) = \psi_{\vec{\lambda}}^{I_J} (q^{h+k}) \ . 
\ee
If we assume this relation, the above expression for the index can be evaluated easily using the completeness of the characters. We get
\begin{align} \label{eq:SchurAA}
\begin{split} 
 \CI_{J^h [k]} (q) &= \frac{1}{\prod_{i=1}^r (q^{d_i}; q)} \oint [d\vec{z}] \PE \left[ - \frac{q^{h+k}}{1-q^{h+k}} \chi_{\textrm{adj}} (\vec{z}) \right] \sum_{\vec\lambda} \chi_{\vec\lambda} (q^{\vec\rho}) \chi_{\vec\lambda} (\vec{z}) \\
 &= \frac{1}{\prod_{i=1}^r (q^{d_i}; q)} \PE \left[ - \frac{q^{h+k}}{1-q^{h+k}} \chi_{\textrm{adj}} (q^{\vec\rho}) \right] \ . 
\end{split}
\end{align}
Here we used the completeness of the character $\sum_{\vec\lambda} \chi_{\vec\lambda} (\vec{z})\chi_{\vec\lambda} (\vec{w}) = \delta(\vec{z}-\vec{w})$. 
The character of the adjoint evaluated at $\vec{z}=q^{\vec\rho}$ is given by
\be
 \chi_{\textrm{adj}} (q^{\vec\rho}) = \sum_{i=1}^{r} \chi_{d_i - 1} (q) = \sum_{i=1}^r [2 d_i -1]_q \ , 
\ee
where $\chi_j$ denotes the character of the spin-$j$ representation of $SU(2)$. 

\paragraph{Example: $A_{N-1}^N [k] = (A_{N-1}, A_{k-1})$ theory}
This theory is engineered by putting 6d $\CN=(2, 0)$ theory of type $J=A_{N-1}$ on a sphere with $I_{N, k} \equiv A_{N-1}^N[k]$ puncture. 
It was conjectured \cite{Cordova:2015nma,Song:2015wta} that the Schur index of this theory is given by the vacuum character of the $W(k, k+N)$ minimal model. The vacuum character for the $W(k, k+N)$ minimal model is given in \cite{Andrews1999a} as 
\begin{align}
\chi_0^{W(k, k+N)}(q) = \left( \frac{(q^{k+N}; q^{k+N})}{(q;q)} \right)^{k-1} \prod_{a=1}^{k-1} (q^{N+a}; q^{k+N})^a (q^a; q^{k+N})^{k-a} \ .
\end{align}
We can rewrite the character using PE as
\begin{align}\label{eq:Wchar}
\begin{split}
\chi_0^{W(k, k+N)}(q) &= \PE \left[ (k-1)\left( \frac{q}{1-q} - \frac{q^{k+N}}{1-q^{k+N}} \right) - \sum_{a=1}^{k-1} \frac{a q^{N+a} + (k-a)q^a }{1-q^{k+N}} \right] \\
&= \PE \left[ \frac{(q-q^k)(q-q^N)}{(1-q)^2 (1-q^{k+N})}\right] \ . 
\end{split}
\end{align}
Note that the character is symmetric under the exchange of $k$ and $N$. 

Now, let us prove that \eqref{eq:SchurAA} indeed reproduces the corresponding character. We have
\begin{align}
\begin{split}
 \CI_{A_{N-1}^N [k]} (q)
 &= \frac{1}{{\prod_{i=2}^N (q^i; q)}} \PE \left[ - \frac{q^{N+k}}{1-q^{N+k}} \chi_{\textrm{adj}} (q^{\vec{\rho}}) \right] \ . 
\end{split}
\end{align}
Note that $\chi_{\textrm{adj}}^{SU(N)} (q^{\vec\rho}) = ([N]_q)^2 - 1$, where $[N]_q = \frac{q^{N/2}-q^{-N/2}}{q^{1/2}-q^{-1/2}}$. Putting this back to the expression above, we obtain
\begin{align}
\begin{split}
 \CI_{A_{N-1}^N [k]} (q) &= \PE \left[ \frac{q^2 + \cdots + q^N}{1-q}- \frac{q^{N+k}}{1-q^{N+k}} \left( \frac{q^{-N} \left(q - q^N\right) \left(1 - q^{N+1}\right)}{(1-q)^2}  \right) \right] \\
  &= \PE \left[ \frac{(1-q^{N-1})(q^2 - q^{k+1})}{(1-q)^2 (1-q^{k+N})} \right] \ . 
\end{split}
\end{align}
This is exactly equal to the expression in \eqref{eq:Wchar}, so the proof is done. 

\paragraph{Example: $D_N^{2N-2}[k]$ theory}
When $J = D_N$, we have
\be
 \chi_{\textrm{adj}} (q^{\vec\rho}) = \frac{q^{1-2N}(1-q^N)(1-q^{2N-1})(q^2+q^N)}{(1-q)(1-q^2)} \ , 
\ee
so that we obtain
\begin{align}
 \CI_{D_N^{2N-2} [k]} (q) &= \PE \left[ \frac{q^2 + q^4 + \cdots q^{2N-2} + q^N}{1-q} - \frac{q^{2N-2+k}}{1-q^{2N-2+k}} \frac{q^{1-2N}(1-q^N)(1-q^{2N-1})(q^2+q^N)}{(1-q)(1-q^2)} \right] \nn \\
 &= \PE \left[ \frac{(1-q^N)(q^2+q^N)(1 - q^{k-1})}{(1-q)(1-q^2)(1-q^{2N-2+k})} \right] \ . 
\end{align}
This is expected to be the closed form formula for the character of the corresponding VOA $\CW({D_N})$ with the central charge given by
\begin{align}
 c_{2d} = - \frac{r(k-1)(h + k +hk)}{h+k} \ ,
\end{align}
where $r=N$, $h=2N-2$. 

\subsection{AD theories of type $(J^{h}[k], Y)$}
\paragraph{$(J^h[k], F)$ theory}
The $(J^h[k], F)$ theory is engineered by adding a full regular puncture in addition to $J^h[k]$ puncture on a sphere. From the TQFT structure, it is straight-forward to write the Schur index once we know the corresponding wave functions for the punctures. The wave function for the irregular puncture $J^h[k]$ is given as in \eqref{eq:IrrSchur}. The wave function for the full regular puncture is given by
\begin{align}
 \psi_{\vec\lambda}(\vec{z}) = \PE \left[ \frac{q}{1-q} \chi_{\textrm{adj}}(\vec{z}) \right] \chi_{\vec\lambda}(\vec{z}) \ . 
\end{align}
Now, combining the above two, we obtain
\begin{align}
\begin{split}
 I_{(J^b[k], F)} &= \sum_{\vec\lambda} \psi_{\vec \lambda}^{J^h[k]}(q)  \psi_{\vec \lambda} (\vec{z}) \\
  &= \oint [d\vec{x}] \PE \left[ - \frac{q^{h+k}}{1-q^{h+k}} \chi_{\textrm{adj}} (\vec{x}) \right] \PE \left[ \frac{q}{1-q} \chi_{\textrm{adj}} (\vec{z}) \right]  \sum_{\vec\lambda} \chi_{\vec\lambda} (\vec{x}) \chi_{\vec\lambda} (\vec{z}) \\
  &= \PE \left[ \frac{q - q^{h+k}}{(1-q)(1-q^{h+k})}\chi^F_{\textrm{adj}} (\vec{z}) \right] \ . 
\end{split}
\end{align}
This is expected to be the character for the VOA given by the affine Lie algebra $J_{-\half k_F}$. We have checked this result against a number of examples as we discussed in the previous section. 
Our conjecture here has been proven recently in \cite{kac2017remark}.

\paragraph{$(J^h[k], Y)$ theory}
From the Jacobson-Morozov theorem every nilpotent orbit are in one-to-one correspondence with the $SU(2)$ embedding.  Therefore, one can label the puncture by a $SU(2)$ embedding $\Lambda$ to $J$. Let us denote $\Lambda_Y$ to be the $SU(2)$ embedding into $J$. When $J=A_{N-1}$, the $SU(2)$ embeddings are labelled by partitions of $N$. One can obtain the theory corresponding to the general puncture by starting with the full regular puncture and performing a partial closure. From the field theory side, this means that we Higgs the $(J^h[k], F)$ theory by giving a nilpotent vev $\Lambda_Y(\s^+)$ to the moment map operator for the flavor group $J$. In terms of the VOA, it can be realized as a quantum Drinfeld-Sokolov reduction by the $SU(2)$ embedding $\Lambda_Y$. 

The TQFT prescriptions for the general regular punctures are studied in various places \cite{Gadde:2011ik,Gadde:2011uv,Lemos:2012ph,Mekareeya:2012tn,Gadde:2013fma}. Let us decompose the adjoint representation of $J$ in terms of the $SU(2)$ embedding $\Lambda_Y$: 
\begin{align} \label{eq:slodowy}
 \textrm{adj} \to \bigoplus_j R_j \otimes V_j \ , 
\end{align}
where $R_j$ is a representation of the commutant of $\Lambda_Y$ in $J$, and the $V_j$ is the spin-$j$ irreducible representation of $SU(2)$. The $\oplus_j R_j$ gives the Slodowy slice of $J$ given by $Y$. 
Now, for the Schur index, the wave function for the regular puncture of type $Y$ is given by
\begin{align}
 \psi^{Y}_{\vec\lambda}(\vec{a}) = \PE \left[\sum_j \frac{q^{j+1}}{1-q} \Tr_{R_j} (\vec{a}) \right] \chi_{\vec\lambda} ( \vec{a}q^{\Lambda_Y} ) \ , 
\end{align}
where $q^{\Lambda_Y}$ is defined as the image of the map $\Lambda_Y: SU(2) \to J$. For example, when $J=SU(9)$ and $\Lambda_Y$ is given by the partition $[3, 2, 2, 1, 1]$, the commutant is given by $S[U(1)U(2)^2]$ and we set $\vec{a}q^{\Lambda_Y} = (a q^1, a q^0, a q^{-1}, b q^{\half}, b q^{-\half}, cq^{\half}, cq^{-\half}, d, e )$ with $abcde = 1$. 

From this, it is easy to compute the Schur index for the $(J^h[k], Y)$ theory. We obtain
\begin{align}
\label{eq:schur1ir1r}
 \CI_{(J^b[k], Y)} &= \sum_{\vec\lambda} \psi_{\vec \lambda}^{J^h[k]}(q) \psi^{Y}_{\vec\lambda}(\vec{a}) \nn \\
  &= \oint [d\vec{x}] \PE \left[ - \frac{q^{h+k}}{1-q^{h+k}} \chi_{\textrm{adj}} (\vec{x})  
  + \sum_j \frac{q^{j+1}}{1-q} \Tr_{R_j} (\vec{a}) \right] \sum_{\vec\lambda} \chi_{\vec\lambda} (\vec{x}) \chi_{\vec\lambda} ( \vec{a}q^{\Lambda_Y} )  \nn \\
  &= \PE \left[\sum_j \frac{q^{j+1}}{1-q} \Tr_{R_j} (\vec{a}) - \frac{q^{h+k}}{1-q^{h+k}} \chi_{\textrm{adj}} (\vec{a}q^{\Lambda_Y}) \right] \ . 
\end{align}
When $Y$ is the null puncture $N$ or the $\Lambda_Y$ is given by the principal embedding, the commutant of $\Lambda_Y$ is empty and $\textrm{adj} \to \oplus_{d_i} V_{d_i}$ where $d_i$ are the degrees of Casimirs. In this case, $q^{\Lambda_Y} = q^{\vec{\rho}}$, so that we reproduce the result of \eqref{eq:SchurAA}. 


\paragraph{Example: $(A_{N-1}^N[k], [N]) = A_{N-1}^N[k]$ theory} Let us consider the example of $(A_{N-1}^N[k], Y) $ with the puncture $Y$ fully closed to a null puncture. The theory should be the same as the theory on a sphere with only one irregular puncture of type $A_{N-1}^N[k] (=I_{N,k})$. 
Following equation \eqref{eq:schur1ir1r}, the Schur index is given by
\begin{equation}
\CI_{(A_{N-1}^N[k],N)}= \PE\left[\sum_{i=2}^{N}\frac{q^i}{1-q}-\frac{q^{N+k}}{1-q^{N+k}}\chi_{\textrm{adj}}^{SU(N)} (q^{\vec{\rho}})\right]=\CI_{A_{N-1}^N[k]},
\end{equation}
which is the same as the Schur index of $A_{N-1}^N[k]$ theory.

\paragraph{Example: $(A_{N-1}^N[k], [N-1,1]) $ theory} Let us consider the example of $(A_{N-1}^N[k], S)=(I_{N, k}, S)$ theory. 
This theory is engineered by putting 6d $\CN=(2, 0)$ theory of type $J = A_{N-1}$ on a sphere with irregular puncture of type $A_{N-1}^N[k]$ (=$I_{N, k}$) and a simple regular puncture. A simple puncture $S$ has $\Lambda_S$ given by the partition  $[N-1,1]$ for $SU(N)$. The commutant is $U(1)$ and $\textrm{adj} \to \oplus_{j=0}^{N-2}V_{j} \oplus ((U(1)_2  \oplus U(1)_{-2}) \otimes V_{N/2 - 1})$. Its Schur index is,
\begin{equation}
\label{eq:simplepunc}
\CI_{(A_{N-1}^N[k],S)}(z) = \PE\left[\sum_{i=1}^{N-1}\frac{q^i}{1-q} +\frac{q^{\frac{N}{2}}}{1-q}(z^2+z^{-2})-\frac{q^{N+k}}{1-q^{N+k}}\chi_{\textrm{adj}}^{SU(N)} (zq^{\frac{N-1}{2}},\cdots,zq^{-\frac{N-1}{2}},z^{-1})\right]. 
\end{equation}

When $k=1$, $(A_{N-1}^N[1], [N-1,1])$ is equivalent with $(A_1, A_{2N-1})$ theory and the corresponding VOA is $\CW^{{-N^2\over N+1}}(A_{N-1}, [N-1,1])$. For example, the Schur index of $(A_{2}^3[1], [N-1,1])=(A_1,A_5)$ can be worked out based on equation \eqref{eq:simplepunc},
\begin{equation}
\begin{split}
&\CI_{(A_1, A_5)}(z) = 1+q+(z^2+z^{-2})q^{\frac{3}{2}} + 3q^2 + (2z^2+2z^{-2})q^{\frac{5}{2} }+ (z^4+5+z^{-4})q^3+\cdots.\\
\end{split}
\end{equation}
The results match perfectly with the vacuum character of $\CW^{{- \frac{9}{4}}}(A_{2}, [2,1])$ \cite{Cordova:2015nma,Buican:2015ina,Song:2015wta}.

\paragraph{Example: $(A_{N-1}^N[k], F) = (I_{N, k}, F)$ theory}
Let us consider the example of $(A_{N-1}^N[k], F)=(I_{N, k}, F)$ theory. 
This theory is engineered by putting 6d $\CN=(2, 0)$ theory of type $J = A_{N-1}$ on a sphere with irregular puncture of type $A_{N-1}^N[k]$ (=$I_{N, k}$) and a full regular puncture \cite{Xie:2012hs,Xie:2013jc}. When $N$ and $k$ are coprime, this theory is conjectured to have the VOA \cite{Beem:2013sza,Beem:2014rza} given by $\widehat{\mathfrak{su}}(N)_{-\frac{N(N-1+k)}{N+k}}$ \cite{Xie:2016evu}. From the TQFT structure, the Schur index is given by
\be
 \CI_{(A_{N-1}^N [k], F)} (\vec{z}) = \PE \left[ \frac{q - q^{N+k}}{(1-q)(1-q^{N+k})} \chi_{\textrm{adj}}^{SU(N)} (\vec{z}) \right] \ .
\ee
For example, when $N=2n+1$ and  $k=-2n+1$, we get 
\be
 \CI_{(A_{2n}^{2n+1}[-2n+1], F)} = \PE \left [ \frac{q}{1-q^2} \chi_{\textrm{adj}}^{SU(2n+1)} (\vec{z}) \right] \ , 
\ee
which proves the conjecture made in \cite{Xie:2016evu}. 

When $N=2$, this theory is identical to $(A_1, D_{2+k})$. Especially, when $k=2n-1$, from the TQFT and wave functions, we obtain
\be
 \CI_{(A_1^2 [2n-1], F)} (z) = \frac{1}{(q z^{\pm 2, 0}; q)} \sum_{m\ge 0} (-1)^m q^{\frac{m(m+1)}{2} (2n+1)} \chi_{2m} (z) \ , 
\ee
which exactly agrees with the vacuum character of $\widehat{\mathfrak{su}}(2)_{-\frac{4n}{2n+1}}$. 

For $N=3$, when $(N, k)=1$, the VOA is given by $\widehat{\mathfrak{su}}(3)_{-\frac{3(k+2)}{k+3}}$ for $k>-2$. When $k=-2$, we get $c_{4d}=0$, which means that the theory is trivial. We find the explicit form of the Schur indices to be
\begin{align}
\CI_{(A_2^3 [-2], F) } &= 1 \\
\CI_{(A_2^3 [-1], F) } &= 1+8 q+36 q^2+128 q^3+394 q^4+1088 q^5+2776 q^6+6656 q^7 +\ldots \ ,  \\
\CI_{(A_2^3 [1], F) } &= 1+8 q+44 q^2+192 q^3+718 q^4+2400 q^5+7352 q^6+20992 q^7 +\ldots  \ , \\
\CI_{(A_2^3 [2], F) } &= 1+8 q+44 q^2+192 q^3+726 q^4+2456 q^5+7640 q^6+22176 q^7 +\ldots  \ , \\
\CI_{(A_2^3 [4], F)} &= 1+8 q+44 q^2+192 q^3+726 q^4+2464 q^5+7704 q^6+22520 q^7 +\ldots \ , 
\end{align}
where we set $z_i=1$ for simplicity. 

\section{Higgs branch and associated variety of VOA} \label{sec:Higgs}

In this section, we study Higgs branches of the generalized Argyres-Douglas theories from the vertex operator algebra (VOA) or chiral algebra perspective. For a given VOA $\CV$, it is possible to construct the associated variety $X_\CV$, which was studied by Arakawa \cite{MR3456698}. It has been conjectured that the associated variety of the VOA agrees with the Higgs branch \cite{BRVOA}. We give some evidence for the case of Argyres-Douglas theories by computing the Hilbert series of the Higgs branch and compare against the one obtained from the TQFT description of the Hall-Littlewood index.  We also use the 3d mirror symmetry to check some of the examples.

\subsection{Higgs branch from the associated variety of VOA}

For a VOA, one can associate a Poisson variety \cite{MR3456698}. The construction goes as follows. Recall that a vertex algebra $\CV$ is a vector space
equipped with an element $\textbf{1} \in \CV$ (or $\ket{0} \in \CV$) called the vacuum, $T \in \textrm{End}(V)$ called the translation operator, and a linear map 
\begin{equation}
Y(\cdot,z):   \CV\rightarrow \textrm{End}(\CV)[[z,z^{-1}]],~~a \mapsto Y(a, z)=a(z)=\sum_{n\in Z} a_{(n)} z^{-n-1}.
\end{equation}
called the state-operator correspondence, such that 
\begin{align}
\begin{split}
& \textbf{1}(z)=id_\CV, \\
& a_{(n)}b=0~\textrm{for}~n \gg 0 \\
& a_{(n)}\textbf{1}=0~\textrm{for}~n\geq 0,~~\textrm{and}~~a_{(-1)}\textbf{1}=a \\
& (Ta)(z)=[T, a(z)]={d\over dz}a(z) \\
& (z-w)^n[a(z), b(w)]=0~\textrm{in}~\textrm{End}(\CV)~\textrm{for}~n\gg0.
\end{split}
\end{align}
There is a filtration $F^{\bullet} \CV$ on any vertex algebra \cite{Li2005}: Set $F^0\CV=\CV$, and 
\begin{equation}
F^p \CV=\textrm{Span} \{a_{(-i-1)}b; a\in \CV,~b\in F^{p-i}\CV,~i\geq 1\} \ . 
\end{equation}
Then we have a decreasing filtration
\begin{equation}
\CV=F^0\CV\supset F^1 \CV \supset F^2 \CV \supset \ldots \ . 
\end{equation}
The subspace $F^1 \CV$ is the linear span of $a_{(-2)} b$ with $a,b \in \CV$. Now, we define 
\begin{equation}
R_\CV=\CV/F^1\CV.
\end{equation} 
This is called the Zhu's $C_2$-algebra of $\CV$ \cite{zhu1996modular}. 
In short, $R_\CV$ is the set of states that are only generated by the $(-1)$-modes of any generators in a VOA. 

$R_\CV$ has a Poisson structure and the Poisson bracket is given by 
\begin{equation}
\{a, b\}=a_{(0)} b~~\textrm{for}~a,b\in R_\CV
\end{equation}
From now on, we assume $R_\CV$ is finitely generated as a ring, and define the associated variety of $\CV$ as
\begin{equation}
X_\CV=\textrm{Spec} R_\CV \ . 
\end{equation}
One can also define associated variety for a (non-vacuum) module of a vertex algebra. 

Now, we conjecture that for the Argyres-Douglas theories, the Higgs branch chiral ring is given by $R_\CV$ and the Higgs branch is given by the associated variety $X_\CV$  \cite{BRVOA}: 
\begin{Conjecture}
Zhu's algebra $R_\CV$ is identical to the Higgs branch chiral ring of the Argyres-Douglas theories. For this case, the Higgs branch is given by the associated variety $X_\CV$. 
\begin{align}
 \CM_{\rm{Higgs}} = X_\CV \ , \qquad \IC[\CM_{\rm{Higgs}}] = R_\CV. 
\end{align}
\end{Conjecture}
For a general 4d $\CN=2$ SCFT, the above relation has to be modified, since $R_\CV$ is in general (conjectured to be) given by the Hall-Littlewood (HL) chiral ring \cite{BRVOA}. The HL chiral ring is generated by the operators contributing to the HL index, which is not in general identical to the Higgs branch operators. But they are indeed identical when the 4d $\CN=2$ SCFT is given by a quiver gauge theory with no loops \cite{Gadde:2011uv}. We conjecture this is also the case for the AD theories.   

\paragraph{Examples}
Let us consider some simple examples. The VOA $(A_1, A_{2n})$ theory is given by a simple Virasoro minimal model. For these theories, $L_{-1}\ket{0}$ is a null state, therefore $R_\CV$ is trivial and $X_\CV$ is a point. 

The VOA for the $(A_1, D_{2n+1})$ theory, is given by $\widehat{\mathfrak{su}}(2)_{-\frac{4n}{2n+1}}$. In this case, we have 3 generators $J^+, J^0, J^-$. So $R_\CV$ is generated by $J^+_{-1}, J^0_{-1}, J^-_{-1}$. But there is a relation $L_{-2} \ket{0} \sim (J^+_{-1} J^-_{-1} +J^0_{-1}J^0_{-1})\ket{0}$ which is coming from the Sugawara construction of the Virasoro generators from the affine Lie algebra. Since $L_{-2} \ket{0} \in F^1 \CV$, it has to be modded out.\footnote{The vertex operator algebra or conformal vertex algebra contains an element called the conformal vector $\omega = L_{-2} \ket{0}$, which is mapped to the stress tensor: $Y(\omega, z) = L(z)$} This gives us
\begin{align}
 R_\CV = \IC[x, y, z]/\langle xy+z^2\rangle \ . 
\end{align}
We see that the associated variety 
\begin{align}
X_\CV = \textrm{Spec}(R_\CV) = \{x, y, z \in \IC ~|~ xy+z^2=0 \} = \IC^2/\IZ_2
\end{align}
is identical to the Higgs branch of the $(A_1, D_{2n+1})$ theory. 

\paragraph{Non-maximal punctures}
Now let us consider the VOA $W^{k_{2d}}(J, Y)$ associated to our AD theories $(J^b[k], Y)$, where we partially close the puncture as the one labeled by $Y$. We 
also assume that the level $k_F$ is admissible, i.e. $b=h$.  The associated variety for this case has been already determined by Arakawa \cite{MR3456698}, and the 
answer takes the following form
\begin{equation}
X_\CV=X_M\cap S_Y.
\end{equation}
Here $X_M$ is the associated variety of the Kac-Moody algebra $\widehat{J}_{k_{2d}}$, and $S_Y$ is the Slodowy slice defined using the nilpotent element of $Y$. 
Recall that the Slodowy slice defined by $Y$ is given by $S_Y = \oplus_j R_j$ under the decomposition $\mathrm{adj} \to \bigoplus_j R_i \otimes V_j$. The space $X_\CV$ is identical to the Higgs branch obtained by the partial closure of the puncture given by $Y$, which is the commutant of the nilpotent element $\Lambda_Y(\s^+)$ inside the Higgs branch $X_\CV$.

The associated variety $X_M$ is given by the closure of a nilpotent orbit $\IO[k_{2d}]$, depending only on the denominator of $k_{2d} = -\half k_F$. 
If the level is given in the following form
\begin{equation}
k_{2d} =-h+{p\over q},~~(p,q)>1, p\geq h, 
\end{equation}
the associated variety is given by
\begin{align}
 X_M = \overline{\IO}_q \ . 
\end{align}
Here the nilpotent orbits $\IO_q$ for each $J$ is given in tables \ref{table:ADorbits}, \ref{table:E6orbits}, \ref{table:E7orbits}, \ref{table:E8orbits}, reproduced from \cite{MR3456698}. 

\subsection{Examples} \label{sec:HiggsEx}
In this subsection, we focus on theories with $J=A_{N-1}$, $b=N$. This gives the corresponding chiral algebra to be the Kac-Moody algebra $\widehat{\mathfrak{su}}(N)_{k_{2d}}$ with level taking the following form
\begin{equation}
k_{2d}=-N+{N\over N+k},~~(k,N)=1.
\end{equation}
If $k>0$, then the associated variety is the principal nilpotent cone of the $A_{N-1}$ Lie algebra. 
This is identical to the Higgs branch of the (3d $\CN=4$) quiver gauge theory 
\begin{align}
 (1) - (2) - (3) - \ldots - (N-1) - [N] \ .
\end{align}
Here, each nodes of $(n)$ means $U(n)$ gauge group and $[N]$ refers to $SU(N)$ flavor group. 

If $k<0$, then the associated variety is the closure of the following nilpotent orbit labelled by a partition of $N$
\begin{equation}
[\underbrace{q,\ldots, q}_r, s],~~s\leq q-1 \ , 
\end{equation}
where $q=N+k$. The nilpotent orbit can be identified with the Higgs branch of a quiver gauge theory associated with the transpose of the partition above, which is $[\underbrace{r+1, \ldots, r+1}_s,\underbrace{r,\ldots, r}_{q-s}, ]$. The quiver is:
\begin{align}
(r)-(2r)-\ldots-((q-s)r)-((q-s)r+r+1)-\ldots-(N-(r+1))-[N].
\end{align}

\paragraph{$\mathbf{k>0}$}
According to Arakawa's result, the Higgs branch of these theories is identified with nilpotent cone of $A_{N-1}$ Lie algebra.

\paragraph{$\mathbf{k=-1}$}
For this case, $k_{2d} = -N + \frac{N}{N-1}$ so that $q=N-1 < h^\vee = N$. The associated variety for the VOA is given by the nilpotent orbit labelled by the partition $(N-1, 1)$. This gives the following quiver described by the transpose of the partition $(2, 1, 1, \ldots, 1)$:
\be
 (1) - (2) - (3) - \ldots - (N-2) - [N]
 \label{kequalminusone}
\ee
The Higgs branch of the above quiver gauge theory is the closure of sub-regular nilpotent orbit. 

\paragraph{$\mathbf{k=-N+2}$}
For this case, $k_{2d} = -N +\frac{N}{2}$. We choose $N=2n+1$ to be an odd number. Then the nilpotent orbit is given by the partition $(2, 2, \ldots, 2, 1)$. The transpose of the partition is given by $(2N+1, N)$, which gives the quiver
\be
 (N) - [2N+1] \ .
 \label{mini}
\ee
The Higgs branch of the above quiver gauge theory is identical to the Higgs branch of the $(A_{N-1}^N[-N+2], F)$ theory. 

\paragraph{Minimal nilpotent orbits}
Let us consider some theories whose corresponding VOAs have non-admissible levels. 
The Higgs branch of the $(D_4^4[-3], F)$, $(E_6^9[-8],F)$, $(E_7^{14}[-13],F)$, $(E_8^{24}[-23],F)$ theories are given by the minimal nilpotent orbit $D_4, E_6, E_7, E_8$. 
These are the same theories as the rank 1 SCFTs with $D_4, E_{6, 7, 8}$ global symmetries.  
Affine vertex operator algebras with non-admissible levels have been studied recently \cite{Arakawa:2016ad, Arakawa:2016aa}, and the corresponding associated variety is exactly the minimal nilpotent orbit. 

\subsection{Check using  3d mirror}
It is possible to write down the 3d mirror quiver $B$ for the class of theory $(J^b[k], Y)$ with 
$J=A_{N-1}, b=N, (k,N)=1$ and $Y=F$ being a full puncture (see section \ref{sec:ADthy}). The Coulomb branch of  
this mirror quiver B (in the IR limit) should give the Higgs branch of the original 4d theory $A$. It is often possible to find another 3d quiver $C$ whose Higgs branch would 
be the same as the IR Coulomb branch of theory $B$. In turn, the Higgs branch of the quiver $C$ coincides with the Higgs branch of original 4d theory $A$. Therefore, the quiver $C$ can be very useful
to check Arakawa's result. We will discuss several examples here, and leave the full check of Arakawa's result to an interested reader. 

Consider a theory given by $\mathbf{J=A_{2N},~b=2N+1,~k=-2N+1}$, and a full regular puncture $Y=F$. 
The 3d mirror of this theory is described by the following quiver \cite{Xie:2016evu} (here we ungauge the $U(1)$ gauge group, see figure \ref{3dmirror}):
\begin{align}
\begin{split}
&(1)-(2)-\ldots-(N)-(N)-\ldots-(2)-(1)\\
&~~ ~~~~~~~~~~~~~~~~~~~~~~| ~~~~~~~~ |    \\
&~~ ~~~~~~~~~~~~~~~~~~~~~[1] ~~~~~~[1] 
\end{split}
\end{align} 
The mirror of the above quiver can be found using the Hanany-Witten construction, and it is exactly as the one shown in \eqref{mini}.
The Higgs branch of this quiver (which is identical to the Higgs branch of the original 4d theory) is exactly given by the closure of nilpotent orbit labelled as $[N+1, N]$. 

Let us now consider the theory with $\bf{J=A_{N-1},~b=N,~k=1}$ and a full regular puncture ($Y=F$). Its 3d mirror is given as below (see also figure \ref{3dmirror}): 
\begin{align}
[N]-(N-1)-\ldots-(2)-(1)
\end{align}
This quiver is self-mirror, and the Coulomb branch is given by the nilpotent cone of the $\mathrm{su}(N+1)$ Lie algebra. This agrees with the associated variety of the corresponding VOA. 

Finally, Let's now consider the theory with $\bf{(J=A_{N-1},~b=N,~k=-1)}$ and a full regular puncture, its 3d mirror is (see figure. \ref{3dmirror} and we ungauge the $U(1)$ gauge group.)
\begin{align}
\begin{split}
&[N-2]-(N-2)-(N-2)-(N-3)-\ldots-(2)-(1)\\
&~~ ~~~~~~~~~~~~~~~~~~~~~~~~~~~~~~|    \\
&~~ ~~~~~~~~~~~~~~~~~~~~~~~~~~~~~[1]
\end{split}
\end{align} 
Using the Hanany-Witten construction, one can find its mirror as shown in \eqref{kequalminusone}. 

\subsection{Check using TQFT}
The Hall-Littlewood limit of the superconformal index is defined as 
\begin{align}
 \CI_{HL}(t) = \Tr_{\CH_{HL}} (-1)^F t^{E-R} \ , 
\end{align}
where the trace is over the states satisfying $E-2R-r=0$ and $j_1=0$. The Hall-Littlewood (HL) index is conjectured to be the same as the Hilbert series on the Higgs branch for a large class of theories.\footnote{This fails for the quiver gauge theory with loops.} We will use the TQFT description to compute the HL indices for the AD theories and verify against the direct computation for the nilpotent orbits for the possible cases. From the TQFT we also derive concise expressions for the Hilbert series of the Higgs branch for certain AD theories. 

\subsubsection{Wave function for the irregular puncture}
We follow similar strategy as in the case of Schur index. First, we come up with the wave function for the Gaiotto state (minimally irregular singularity) realizing the pure Yang-Mills (YM) theory. Second, we try to guess the wave functions for other irregular singularities. 

The Hall-Littlewood index for the pure YM theory with gauge group $J$ is given by
\be
 \CI_{HL}^J = \oint [d\vec{z}] I_{\textrm{vec}} (\vec{z}) = \oint [d\vec{z}] \PE \left[ -t \chi_{\textrm{adj}} (\vec{z}) \right] = (1-t)^r \oint [d\vec{z}] \prod_{\vec{\a} \in \Delta_J} (1-t \vec{z}^{\vec\alpha}) \ . 
\ee
We find a closed form expression for the $G=SU(N)$ case as
\be
 \CI_{HL}^{SU(N)} = \prod_{n=1}^{N-1} (1 - t^{2n+1}) \ . 
\ee
From the TQFT structure of the index, we should be able to obtain the index from the sphere with two irregular punctures of the same type $I_J \equiv I_J[-h+1]$. Therefore, we write the index as
\be
 \CI_{HL}^{J} = \sum_{\vec\lambda} \psi_{\vec\lambda}^{I_{J}} \psi_{\vec\lambda}^{I_{J}} \ , 
\ee 
where the wave function for the $I_{J}$ puncture is given by
\be
\psi_{\vec\lambda}^{I_{J}} =\oint [d\vec{z}] \PE \left[  -t \chi_{\textrm{adj}} (\vec{z}) \right] \psi_{\vec\lambda}(\vec{z}) = \oint [d\vec{z}] P^{HL}_{\vec\lambda} (\vec{z}) \ .  
\ee
And the wave function for the full regular puncture is given as
\begin{align}
\psi_{\vec\lambda}(\vec{z}) = \PE \left[ t \chi_{\textrm{adj}} (\vec{z}) \right] P^{HL}_{\vec\lambda}(\vec{z}) \ , 
\end{align}
where $P^{HL}_{\vec\lambda}(\vec z)$ is the (normalized) Hall-Littlewood polynomial. We have chosen our normalization of the Hall-Littlewood polynomial so that
\begin{align}
 \oint [d\vec{z}] I_{\textrm{vec}}(\vec{z}) \psi_{\vec{\lambda}}(\vec{z}) \psi_{\vec{\mu}}(\vec{z}) 
 =  \oint [d\vec{z}] \PE \left[ t \chi_{\textrm{adj}} (\vec{z}) \right] P^{HL}_{\vec\lambda}(\vec{z}) P^{HL}_{\vec\mu}(\vec{z})
 = \delta_{\vec\lambda \vec\mu} \ . 
\end{align}
Using the orthonormality of the Hall-Littlewood polynomial under the vector multiplet measure, we get
\begin{align}
 \sum_{\vec \lambda} \psi_{\vec\lambda}^{I_J} \psi_{\vec\lambda}^{I_J} = \oint [d\vec{z}][d\vec{z'}] P^{HL}_{\vec\lambda}(\vec{z}) P^{HL}_{\vec\lambda}(\vec{z'}) 
  = \oint [d\vec{z}]\PE \left[ -t \chi_{\textrm{adj}} (\vec{z}) \right] = \CI^J_{HL} \ ,
\end{align}
so that our wave function for the irregular puncture $I_J$ is indeed consistent. 

Let us write some explicit examples. For $J=A_1$, we get ($I_{N, k} \equiv A_{N-1}^N[k]$)
\begin{align}
 \psi_{\lambda}^{I_{2, -1}} = 
 \begin{cases}
 	\sqrt{1-t^2} & (\lambda = 0) \ , \\
	-t \sqrt{1-t} & (\lambda =  2) \ .
 \end{cases}
\end{align}
When $G=A_2$, we obtain
\begin{align} 
\begin{split}
 \psi_{(\lambda_1, \lambda_2)}^{I_{3, -2}} = 
\begin{cases}
 	\sqrt{(1-t^2)(1-t^3)} & (\lambda_1, \lambda_2) = (0, 0) , \\
	-t (1-t^2) & (\lambda_1, \lambda_2) = (1, 1), \\
	-t^3 (1-t) & (\lambda_1, \lambda_2) = (2, 2), \\
	t^2 \sqrt{(1-t)(1-t^2)} & (\lambda_1, \lambda_2) = (3, 0), (0, 3), \\
	0 & \textrm{otherwise} \ . 
 \end{cases} 
\end{split}
\end{align}

We conjecture the wave function for $k > 0$ to be
\begin{align} \label{eq:HLpsi0}
\psi_{\vec\lambda}^{J^b[k]} = \prod_{i=1}^{r} (1-t^{d_i})^{\half} \delta_{\vec{\lambda}, \vec{0}} \ , 
\end{align}
where $r=\textrm{rank}(J)$ and $d_i$ are the degrees of the Casimirs of $J$. 
When $k=-1$, we find the wave function to be
\begin{align} \label{eq:HLpsim1}
 \psi_{\vec{\lambda}}^{J^b[k]} = 
 \begin{cases}
 	\prod_{i=1}^r (1-t^{d_i})^{\half} & \vec{\lambda} = \vec{0} \ , \\
	-t^{h - 1}(1-t) \prod_{i=1}^{r-2} (1-t^{d_i})^\half & \vec{\lambda} = \textrm{adjoint}  \ . 
 \end{cases} 
\end{align}
We have checked that the above expressions to be consistent with our previous results and known results. 
It is not obvious to us how to obtain the wave function (and thereby the index) for the general value of $k$.

\subsubsection{AD theory of type $(J^h[k], Y)$}

From the TQFT structure, the Hall-Littlewood index for the $(J^b [k], F)$ theory can be written as
\be
 \CI_{(J^b [k], F)} = \sum_{\vec \lambda} \psi_{\vec \lambda}^{J^b [k]} \psi_{\vec \lambda} (\vec{z}) \ .
\ee
Plugging in the expression \eqref{eq:HLpsi0} to above, it is straight-forward to obtain (for $k>0$)
\be \label{eq:HLidxA}
 \CI_{(J^b [k], F)} = \frac{\prod_{i=1}^r (1-t^{d_i}) }{(1-t)^r \prod_{\vec{\a} \in \Delta_{J}} (1 - t \vec{z}^{\vec{\a}})} \ , 
\ee
where we used the fact that $P^{HL}_{\vec{\lambda} = \vec{0}} (\vec{z}) = \prod_{i=1}^r (1-t^{d_i})^\half $. 
Here $r = \textrm{rank}(J)$ and $d_i$ are the degrees of the Casimirs of $J$. Note that there is no dependence on $k$, which reflects the fact that the Higgs branch for each $k>0$ is the same. This agrees with the result of \cite{MR3456698} that the associated variety of the VOA for $k>0$ is always given by a principal nilpotent cone. We claim that this expression is the closed-form expression for the Higgs branch (and equivalently the Coulomb branch since it is self-mirror) of the 3d $\CN=4$ theory $T[J]$ of \cite{Gaiotto:2008ak} with $J \in ADE$. This result also agrees with that of \cite{Cremonesi:2014kwa}.  

\paragraph{Example: $(A_{N-1}^N[k], F) = (I_{N, k}, F)$ theory}
From the TQFT structure, the Hall-Littlewood index for the $(A_{N-1}^N[k], F) = (I_{N, k}, F)$ theory for $k>0$ can be written as 
\be \label{eq:HLidxA}
\CI_{(A_{N-1}^N[k], F)} = \PE \left[ t \chi_{\textrm{adj}} (\vec{z}) \right] \prod_{n=2}^N (1-t^n) = \frac{\prod_{n=2}^N (1-t^n)}{(1-t)^{N-1}  \prod_{\vec\alpha \in \Delta_{SU(N)}} (1-t \vec{z}^{\vec\alpha})} \ . 
\ee
This index is indeed the same as the Hilbert series of the Higgs branch of $T[SU(N)]$ theory. This is also the same as the Hilbert series of the Coulomb branch because the $T[SU(N)]$ theory is self-mirror.

\paragraph{Example: Non-principal orbits}
Let us consider when the associated variety is given in terms of a non-principal orbit. For $k=-1$, one can obtain the Hall-Littlewood index using \eqref{eq:HLpsim1}. We find that when $J$ is of $A$ or $D$ type, the index agrees with the Hilbert series of the sub-regular nilpotent orbit \cite{Hanany:2016gbz}. 

For example, consider the $(A_2^3[-1], F) = (A_2, A_2) = (A_1, D_4)$ theory. The corresponding VOA is given by $\widehat{\mathfrak{su}}(3)_{-\frac{3}{2}}$. The wave function for $I_{3, -1}$ is given by
\be
 \psi_{(\lambda_1, \lambda_2)}^{I_{3, -1}} = 
 \begin{cases}
  \sqrt{(1-t^2)(1-t^3)} & (\lambda_1, \lambda_2) = (0, 0) \ , \\
  - t^2 (1-t) & (\lambda_1, \lambda_2) = (1, 1) \ . 
 \end{cases}
\ee
It reproduces the correct Hall-Littlewood index of the $(A_2, A_2)=(A_1, D_4)$ theory. We know that the Higgs branch of this theory is given by the minimal nilpotent orbit of $SU(3)$, which is identical to the 1-instanton moduli space of $SU(3)$. 

\paragraph{Non-maximal punctures}
As in the Schur case, it is straight-forward to obtain the theory for the general regular puncture $Y$ via nilpotent Higgsing. The adjoint representation of $J$ is now decomposed in terms of the commutant of the $SU(2)$ embedding $\Lambda_Y$ as $\textrm{adj} \to \bigoplus_j R_j \otimes V_j$. Now, the wave function for the regular puncture of type $Y$ is given by a concise closed form as
\begin{align}
 \psi^Y_{\vec \lambda} (\vec{a}) = \PE \left[ \sum_j t^{j+1} \tr_{R_j} (\vec{a}) \right] P^{HL}_{\vec\lambda} (\vec{a}t^{\Lambda_Y}) \ , 
\end{align}
where various symbols are the same as the ones defined around \eqref{eq:slodowy}. Therefore, the index for the $(J^b[k], Y)$ theory with $k>0$ is given by
\begin{align}
 \CI_{(J^b[k], Y)} = \PE \left[ \sum_j t^{j+1} \tr_{R_j} (\vec{a}) - \sum_{i=1}^r t^{d_i}\right] = \frac{\prod_{i=1}^r (1-t^{d_i})}{\prod_j \prod_{\vec{w} \in R_j} (1-t^{j+1} \vec{a}^{\vec{w}})} \ . 
\end{align}
Here $\vec{w} \in R_j$ denotes the weight vectors of the representation $R_j$. This formula agrees with the one obtained in \cite{Cremonesi:2014kwa}. 

\section{Conclusion} \label{sec:Conclusion}
We systematically studied the chiral algebra/vertex operator algebra corresponding to a large class of Argyres-Douglas theories labeled as $(J^b[k], Y)$. Here $J^b[k]$ denotes an irregular singularity carrying no mass parameters, and $Y$ represents a regular singularity. We restricted to the case where there is no exactly marginal deformations. In particular, all the theories of the form $(A_1, \G)$ with $\G \in ADE$ belongs to this class, so that we recover the result of \cite{Cordova:2015nma,Creutzig:2017qyf}.  
We found a surprisingly simple formula for the Schur index (which is equal to the vacuum character of the corresponding chiral algebra) when $b=h$ utilizing the TQFT description of the index. The Higgs branch of a theory in this class is identified with the associated variety of the corresponding VOA, and we also found a concise formula for its Hilbert series or the Hall-Littlewood index. 

Various nilpotent orbits appear as the Higgs branch of an AD theory. However, not all of them appear. For the $su(N)$ Lie algebra, only the nilpotent orbits of the form $[q,\ldots, q, s]$ appear. It would be interesting to find AD theories whose Higgs branch are given by other nilpotent orbits, or find some obstructions possibly along the line of \cite{Shimizu:2017kzs}.

AD theories in our list can be used to construct various new SCFTs by gauging the flavor symmetry whenever available (they are called the AD matter in \cite{Xie:2017vaf}). The Schur indices of these general SCFTs can be found using the Schur index of the AD matter. These indices can be used to check the S-duality of the Argyres-Douglas theory proposed in \cite{Buican:2014hfa, Xie:2017vaf}. However, not all of the AD theories can be constructed from the gluing of the AD theories considered in this paper, and it is interesting to find the indices of other AD matters used in \cite{Xie:2017vaf}. 

We derived many results for the theory $(J^b[k], Y)$ when $b=h$, where the level of the corresponding chiral algebra is admissible. However, our results also suggests 
that similar results should also hold for $b<h$ where the level for the corresponding VOA is not admissible. It would be interesting to study them further.  
Also, it would be interesting to find more refined partition functions, such as the Macdonald index, Lens space index and the full superconformal index for the AD theories in our class. 

\acknowledgments 
We thank Chris Beem and Leonardo Rastelli for discussions and correspondence. JS and WY would like to thank the organizers of the workshop ``Exact Operator Algebras in Superconformal Field Theories" and Perimeter Institute for hospitality. JS also thanks Kavli IPMU for hospitality where this paper is finalized. 
The work of JS is supported in part by the US Department of Energy under UCSD's contract de-sc0009919 and also by Hwa-Ahm foundation.
The work of DX and WY is supported by Center for Mathematical Sciences and Applications at Harvard University.

\appendix
\section{Some facts about the $\mathcal{N}=2$ index} \label{app:n2idx}
The generic representation of a 4d $\mathcal{N}=2$ SCFT is labeled by the states $|E, R, r, j_1, j_2\rangle$. Here $E$ is the scaling dimension, $R$ is 
the $SU(2)_R$ quantum number, $r$ is the $U(1)_R$ quantum number, and $j_1, j_2$ are left and right spins. 

The superconformal index for a $\mathcal{N}=2$ SCFT is a refined Witten index on $S^3\times S^1$,
\begin{align}
\mathcal{I} = \Tr (-1)^F p^{\frac{1}{2}(E+2j_1-2R-r)} q^{\frac{1}{2}(E-2j_1-2R-r)} t^{R+r},
\end{align}
where the trace is over the BPS states satisfying the condition
\begin{align} \label{eq:IdxQ}
E-2j_2-2R+r = 0,
\end{align}
which are annihilated by a supercharge $\tilde{Q}_{1\dot{-}}$. 
Various limits of the superconformal index exist. Each limits of the index gets contributions from different type of superconformal multiplets. 
In this work, we are interested in the following two limits.

The first is the Schur limit $q=t$ with $p$ arbitrary. The result is independent of the fugacity $p$ automatically. In the trace formula, the Schur index can be written as,
\begin{align}
\mathcal{I}_{\mathrm{Schur}} = \Tr_{\mathrm{Schur}} (-1)^F q^{E-R},
\end{align}
which trace over the BPS states satisfying an additional condition
\begin{equation}
E+2j_1-2R-r = 0,
\end{equation}
besides \eqref{eq:IdxQ} which is used to define the index. They are annihilated by $Q_{1+}$ and $\tilde{Q}_{1\dot{-}}$ supercharges. The Schur index counts the Higgs branch operators, stress tensor and other multiplets, that are in the corresponding vertex operator algebra. 

The second is the Hall-Littlewood limit $q,\,p\rightarrow 0$ with $t$ fixed. The trace formula is,
\begin{equation}
\mathcal{I}_{HL} = \Tr_{HL} (-1)^F t^{E-R}.
\end{equation}
The trace is restricted to states with
\begin{equation}
j_1 = 0, \,\,\,j_2 = r,\,\,\, E=2R+r,
\end{equation}
and are annihilated by supercharges $Q_{1+}$, $Q_{1-}$ and $\tilde{Q}_{1\dot{-}}$.

\section{Nilpotent orbits in Lie algebras}
Here we list the orbits in Lie algebras that appear as an associated variety for the VOA \cite{MR3456698}. 
For the classical type Lie algebra $A$ and $D$, the we use the label from which we can associate to a quiver. The orbit is realized as the Higgs branch of the quiver. See section \ref{sec:HiggsEx} for more detail.
\begin{table}[h]
\begin{center}
  \begin{tabular}{ |c|c|c|}
    \hline
      $g$ & $q$ &$\IO_q$  \\ \hline
           $sl_{l+1}$& any& $[q,\ldots, q, s]$, $0\leq s\leq q-1$ \\ \hline
           $so_{2l}$& odd&$[\underbrace{q,\ldots, q}_{odd}, s]$, ~$0\leq s\leq q$, $s$ odd \\ \hline
           ~&~&$[\underbrace{q,\ldots, q}_{even}, s,1]$, ~$0\leq s\leq q-1$, $s$ odd \\ \hline
           ~&even&$[q+1,\underbrace{q,\ldots, q}_{even}, s]$,~$0\leq s\leq q-1$, $s$ odd \\ \hline
             ~&~&$~q+1,\underbrace{q,\ldots, q}_{even}, q-1,s,1]$,~$0\leq s\leq q-1$, $s$ odd \\ \hline

           \end{tabular}
  \end{center}
  \caption{Orbits in $A$ and $D$ type Lie algebras. }
  \label{table:ADorbits}
\end{table}
For the orbits in the exceptional Lie algebras, we use the Bala-Carter label to denote them. 
\begin{table}[h]
\begin{center}
  \begin{tabular}{ |c|c|c|}
    \hline
      $q$ &$\IO_q$ &$c({p\over q})$  \\ \hline 
      $q\geq 12$& $E_6$ & ${-6(12p-13q)(13p-12q)\over pq}$ \\ \hline
      $9,10,11$& $E_6(a_1)$ & $-{8 (9p-13q)(7p-9q)\over pq}$ \\ \hline
      8& $D_5$ & $-45+1162-{7488\over p}$ \\ \hline
      $6,7$ & $E_6(a_3)$ & $-{36 (6p-13q)(p-2q)\over pq}$ \\ \hline
      $5$ & $A_4+A_1$ & $-2(7p-90)(9p-130)\over 5p$ \\ \hline
      $4$ & $D_4 (a_1)$ & $-18p+524-{3744\over p}$ \\ \hline
      $3$ & $2A_2+A_1$ & $-{18 (p-13)(p-12)\over p}$ \\ \hline
      $2$ & $3A_1$ & $-9p+263-{1872\over p}$ \\ \hline
      $1$ & $0$ & ${78 (p-12)\over p}$ \\ \hline
           \end{tabular}
  \end{center}
  \caption{Orbits in $E_6$ Lie algebra. }
  \label{table:E6orbits}
\end{table}
\begin{table}[h]
\begin{center}
  \begin{tabular}{ |c|c|c|}
    \hline
      $q$ &$\IO_q$ &$c({p\over q})$  \\ \hline 
      $q\geq 18$& $E_7$ & ${-7(18p-19q)(19p-18q)\over pq}$ \\ \hline
      $14,15,16,17$ & $E_7(a_1)$ & ${-9(14p-19 q)(11p-14q) \over pq}$ \\ \hline
      $12,13$ & $E_7(a_2)$ & ${-(106-171 p)(9p-14q)\over pq}$ \\ \hline
      $10,11$ & $E_7(a_3)$ & ${-666 p\over q}+2251-{2394 q\over p} $ \\ \hline
      $9$ & $ E_6(a_1)$ & $56p+2199-{21546\over p}$ \\ \hline
      $8$ & $E_7(a_4)$ & $-{189 p\over 4}+1901-{19152\over p}$ \\ \hline
      $7$ &$A_6$ & $-{(p-18)(48p-931)\over p}$ \\ \hline
      $6$ & $E_7(a_5)$ & $-{3(p-19)(13p-252)\over p}$ \\ \hline
      $5$ & $A_4+A_2$ & ${9\over 5}(-16p+655-{6650\over p})$ \\ \hline
      $4$ & $A_3+A_2+A_1$ & $-{3 (3p-56)(5p-114)\over 2p}$ \\ \hline
      $3$ & $2A_2+A_1$ & $ -18p+721-{7182\over p}$ \\ \hline
      $2$ & $4A_1$ & $-{12(p-21)(p-19)\over p}$ \\ \hline
      $1$ & $0$ & ${133(p-18)\over p}$ \\ \hline
           \end{tabular}
  \end{center}
  \caption{Orbits in $E_7$ Lie algebra.}
  \label{table:E7orbits}
\end{table}
\begin{table}[h]
\begin{center}
  \begin{tabular}{ |c|c|c|}
    \hline
      $q$ &$\IO_q$ &$c({p\over q})$  \\ \hline 
      $q\geq 30$& $E_8$ & ${-8(30p-31q)(31p-30q)\over pq}$ \\ \hline
      $24,25,26,27,28,29$& $E_8(a_1)$ & $-{10(24p-31q)(19p-24q)\over pq}$ \\ \hline
      $20,21,22,23 $ & $E_8(a_2)$ & $-12 {(20p-31q)(13p-20q)\over pq}$ \\ \hline 
      $18,19$ & $E_8(a_3)$ & $-{2400 p\over q}+ 8438-{7440 q\over p}$ \\ \hline
      $15,16,17$ & $E_8(a_4)$ & $-{16(15p-31 q) (7p-15q)\over pq}$ \\ \hline
      $14$ & $E_8(b_4)$ & $-{696p\over 7}+6426-{104160\over p}$ \\ \hline
      $12,13$ & $E_8(a_5)$ & $-{4(23p-60q)(12p-31q)\over pq}$ \\ \hline
      $10,11$ & $E_8(a_6)$ & $-{24(10p-31q)(3p-10q)\over pq}$ \\ \hline
      $9$ & $E_8(b_6)$ & $-{4(4p-135)(11p-372)\over 3p}$ \\ \hline
      $8$ & $A_7$ & $-{3(p-31)(21p-640)\over p}$  \\ \hline
      $7$ & $A_6+A_1$ & $-{342p\over7}+3192-{52080\over p} $ \\ \hline
      $6$ & $E_8(a_7)$ & $-{40(p-36)(p-31)\over p}$ \\ \hline
      $5$ & $A_4+A_3$ & $-{12 (p-31)(3p-100)\over p}$ \\ \hline
      $4$ & $2A_3$ & $-{30 (p-32)(p-31)\over p}$ \\ \hline
      $3$ & $2A_2+2A_1$ & $-{20(p-36)(p-31)\over p}$ \\ \hline
      $2$ & $4A_1$ & $-{12 (p-40)(p-31)\over p}$ \\ \hline
      $1$ & $0$ &$ 248 {(p-30)\over p}$ \\ \hline
           \end{tabular}
  \end{center}
  \caption{Orbits in $E_8$ Lie algebra.}
  \label{table:E8orbits}
\end{table}

\newpage
\bibliographystyle{jhep}
\bibliography{ADhigher}

\end{document}